\documentclass[prd,twocolumn,showpacs,floatfix,amsmath,amssymb,nofootinbib]{revtex4}
\usepackage{graphicx,color,dcolumn,booktabs,bm}
\usepackage{longtable,lscape,comment,braket}
\usepackage{txfonts}
\usepackage{overpic}
\usepackage{amssymb,tipa}
\usepackage{indentfirst}
\usepackage{feynmf}   %{feynmp}
\usepackage{slashed}  %for Feynman symbols
\usepackage{cases}
\usepackage{epstopdf}
\usepackage{psfrag}
\usepackage{subfigure}
\usepackage{color}
\usepackage{multirow}
\graphicspath{{Figures/}} %
\usepackage[colorlinks, citecolor=blue,anchorcolor=red,menucolor=red, linkcolor=red,filecolor=red,runcolor=red,urlcolor=blue,frenchlinks=red]{hyperref}

\usepackage{ulem}
\def\be{\begin{equation}}
\def\ee{\end{equation}}
\def\Tr{\textnormal{Tr}}

\allowdisplaybreaks

\begin{document}
%\begin{CJK}{GBK}{}
\title{Radiative transitions and magnetic moments of the charmed and bottom vector mesons in chiral perturbation theory}
\author{Bo Wang$^{1,2}$}\email{bo-wang@pku.edu.cn}
\author{Bin Yang$^{1}$}\email{bin\_yang@pku.edu.cn}
\author{Lu Meng$^{1}$}\email{lmeng@pku.edu.cn}
\author{Shi-Lin Zhu$^{1,2,3}$}\email{zhusl@pku.edu.cn}
\affiliation{
$^1$ School of Physics and State Key Laboratory of Nuclear Physics and Technology, Peking University, Beijing 100871, China\\
$^2$Center of High Energy Physics, Peking University, Beijing 100871, China\\
$^3$Collaborative Innovation Center of Quantum Matter, Beijing
100871, China}

\begin{abstract}
In this work, we systematically study the radiative decays and
magnetic moments of the charmed and bottom vector mesons with chiral
perturbation theory up to one-loop level. We present the results in
SU(2) and SU(3) cases with the mass splitting in loop diagrams kept
and unkept, respectively. The obtained decay rates for $D^\ast$ and
$B^\ast$ mesons in SU(3) case with the mass splitting kept are:
$\Gamma_{\bar{D}^{\ast 0}\to \bar{D}^0\gamma}=16.2^{+6.5}_{-6.0}$
keV, $\Gamma_{D^{\ast-}\to D^-\gamma}=0.73^{+0.7}_{-0.3}$ keV,
$\Gamma_{D_s^{\ast-}\to D_s^-\gamma}= 0.32^{+0.3}_{-0.3}$ keV,
 and $\Gamma_{B^{\ast+}\to B^+\gamma}=0.58^{+0.2}_{-0.2}$ keV,
$\Gamma_{B^{\ast0}\to B^0\gamma}=0.23^{+0.06}_{-0.06}$ keV,
$\Gamma_{B_s^{\ast0}\to B_s^0\gamma}=0.04^{+0.03}_{-0.03}$ keV. The decay width for
$D^{\ast-}\to D^-\gamma$ is consistent with the experimental
measurement. As a byproduct, the full widths of $\bar{D}^{\ast0}$
and $D_s^{\ast-}$ are
$\Gamma_{\mathrm{tot}}(\bar{D}^{\ast0})\simeq77.7^{+26.7}_{-20.5}~\mathrm{keV}$
and $
\Gamma_{\mathrm{tot}}(D_s^{\ast-})\simeq0.62^{+0.45}_{-0.50}~\mathrm{keV}$,
respectively. We also calculate the magnetic moments of the heavy
vector mesons. The analytical chiral expressions derived in our work
shall be helpful for the extrapolations of lattice QCD simulations
in the coming future.
\end{abstract}
 \maketitle

\section{introduction}\label{Sec1}

Electromagnetic form factors play a very important role in mapping
out the internal structures of nucleons, which offer valuable
information about the distribution of the constituent quarks and
gluon degree of freedom in
nucleons~\cite{Gao:2003ag,Arrington:2006zm,Pacetti:2015iqa,Chupp:2017rkp}.
Probing the shape and inner structure of hadrons still remain an
intriguing and challenging topic. Especially, in recent decades, a
large number of exotic states were observed in experiments, many of
which cannot be readily reconciled with the predictions of the
conventional quark
models~\cite{Chen:2016qju,Guo:2017jvc,Liu:2019zoy}.

Magnetic moments can be related to the form factors by extrapolating
the form factor $\mathcal{G}_M(q^2)$ to zero moment
transfer~\cite{Peskin:1995ev}. Unlike proton and neutron, vast
majority of hadronic states are unstable against strong
interactions~\cite{Tanabashi:2018oca}. Thus, their magnetic moments
cannot be directly measured with the conventional ways due to their
very short lifetime. Therefore, the radiative transition becomes a
very effective way to help us catch a glimpse of quark dynamics in
the hadrons. In addition, the quark model cannot give the nonanalytic
dependence of the magnetic moments, such as the $\log X$ term. These
terms are much more difficult to naively estimate and may be sometimes
singular to give the much enhanced contributions which cannot be
predicted accurately unless carefully calculated.

In this work, we focus on the charmed and bottom vector mesons,
i.e., ($\bar{D}^{\ast0}$, $D^{\ast-}$, $D_s^{\ast-}$), and
($B^{\ast+}$, $B^{\ast0}$, $B_s^{\ast0}$). As a consequence of heavy
quark spin symmetry, the mass shifts between these spin triplets and
singlets are generally small. Because of the small phase space, the
dominant decay channels are one-pion emission transitions and
radiative decays for the charmed vector mesons, while only radiative
decays are allowed for the bottom vector mesons.

From Review of Particle Physics (RPP)~\cite{Tanabashi:2018oca}, only
the width of $D^{\ast\pm}\to D^\pm\gamma$ is known by combining the
decay branching ratio and the total width of $D^{\ast\pm}$. For the
other radiative decay modes, only the branching ratios are
available, and the absolute widths are still absent in experiments.
Even worse, there is no experimental information for the radiative
transitions of the $B^\ast$ mesons.

Many theoretical methods have been applied to study the radiative
decays of the $D^\ast$ and $B^\ast$ mesons, such as various quark
models~\cite{Sucipto:1987qj,Kamal:1992uv,Ivanov:1994ji,Jaus:1996np,Priyadarsini:2016tiu,Choi:2007se},
heavy quark effective theory and vector meson dominance
model~\cite{Colangelo:1993zq}, quark-potential
models~\cite{Godfrey:1985xj,Simonis:2016pnh,Ebert:2002xz,Bose:1980vy,Goity:2000dk,Simonis:2018rld},
QCD sum rules~\cite{Dosch:1995kw,Zhu:1996qy,Aliev:1994nq}, lattice
QCD simulations~\cite{Becirevic:2009xp}, constituent quark-meson
model~\cite{Deandrea:1998uz}, chiral effective field
theory~\cite{Cho:1992nt,Cheng:1992xi,Amundson:1992yp,Casalbuoni:1996pg},
extended Nambu-Jona-Lasinio model~\cite{Deng:2013uca,Luan:2015goa},
and so on.

Here, we adopt the $\mathrm{SU}(3)$ chiral perturbation theory
($\chi$PT) to investigate the radiative decay properties and
magnetic moments of the $D^\ast$ and $B^\ast$ mesons. The framework
of $\chi$PT has been widely used to study the radiative decays and
magnetic moments of the charmed and bottom vector
mesons\footnote{In Refs. \cite{Cho:1992nt,Cheng:1992xi}, Cho {\it et al}
and Cheng {\it et al} calculated the decay widths of $D^\ast\to D\gamma$ and $B^\ast\to B\gamma$
at the tree level in the heavy hadron chiral theory, respectively. Our Lagrangians
are the same with Refs. \cite{Cho:1992nt,Cheng:1992xi} at the leading order. In
Ref. \cite{Amundson:1992yp}, Amundson {\it et al} investigated the same process with the same framework to the next-to-leading order.
But the heavy quark spin symmetry breaking effect is ignored.}~\cite{Cho:1992nt,Cheng:1992xi,Amundson:1992yp,Casalbuoni:1996pg},
the  octet baryons~\cite{Jenkins:1992pi,Meissner:1997hn}, the doubly
charmed and bottom heavy baryons~\cite{Li:2017cfz,Li:2017pxa,Meng:2017dni,Blin:2018pmj}, the
singly heavy baryons~\cite{Jiang:2015xqa,Wang:2018xoc,Meng:2018gan,Wang:2018cre},
as well as the related chiral quark-soliton model for singly heavy
baryons \cite{Yang:2018uoj,Kim:2018nqf}. In our calculations, we
construct the effective Lagrangians with chiral symmetry and heavy
quark symmetry up to $\mathcal{O}(p^4)$. There are two independent
low-energy constants (LECs) at the leading order, which correspond
to the contributions from the light quark and heavy quark
electromagnetic currents, respectively. These two LECs can be
estimated with quark model or other theoretical methods. The
contributions from the tree diagrams at the next-to-leading order
can be absorbed into the ones from the leading order. At the
next-to-next-to leading order, the tree diagram incorporates three
independent LECs, which cannot be determined due to lack of
experimental data. We present our numerical results up to
$\mathcal{O}(p^4)$, and consider the contributions from
$\mathcal{O}(p^4)$ tree diagrams as errors.

Our numerical results are calculated both in SU(2) and SU(3) cases
with the mass splitting kept and unkept in loop diagrams. The
partial decay widths of $D^{\ast-}\to D^-\gamma$ predicted in
different scenarios are consistent with the experimental data.

This paper is organized as follows. The definitions of the
electromagnetic form factors and magnetic moments are given in
Sec.~\ref{Sec2}. The effective Lagrangians are constructed in
Sec.~\ref{sec3}. The analytical expressions and numerical results
for the transition magnetic moments and magnetic moments are
presented in Sec.~\ref{sec4} and Sec.~\ref{Mm}, respectively. A
summary is given in Sec.~\ref{sec5}. Some supplemental materials for
the $B^\ast$ mesons, the loop integrals and an estimation of the
light quark mass with the vector meson dominance model are collected in
Appendices \ref{appdixA}, \ref{appdixB} and \ref{appdixC}, respectively.

\section{electromagnetic form factors and magnetic moments}\label{Sec2}

We first consider the radiative transition process $V\to P\gamma$,
where $V$ stands for the vector mesons ($D^\ast$ or $B^\ast$), and
$P$ denotes the pseudoscalar mesons ($D$ or $B$). The M1 transition
form factor $\mu^\prime(q^2)$ can be defined through a covariant
expression of the hadronic matrix elements~\cite{Casalbuoni:1996pg},
\begin{equation}\label{FormPV}
\langle P(p^\prime)|J_{em}^\mu(q^2)|V(p,\varepsilon_V)\rangle=e
\mu^\prime(q^2) \epsilon^{\mu\nu\alpha\beta} p_\nu q_\alpha
\varepsilon_{V\beta},
\end{equation}
where $J_{em}^\mu$ is the electromagnetic current at hadronic level.
$q_\alpha=(p-p^\prime)_\alpha$ is the transferred momentum, and
$\varepsilon_{V\beta}$ denotes the polarization vector of the
initial vector meson.

The interaction Hamiltonian can then be written as
\begin{equation}\label{HintAJ}
H_{\mathrm{int}}=\int d^3x eA_\mu J^\mu_{em},
\end{equation}
where $A_\mu$ is the photon field.

For a heavy meson $M$ that is composed of a heavy antiquark
$\bar{Q}$ and a light quark $q$, the ground spin doublet
$(P,P^\ast)$ can be represented by a $4\times4$ Dirac-type matrix
$\mathcal{H}$. We use the $\mathcal{H}(p)$ and $\mathcal{H}(v)$ to
denote the heavy meson fields in relativistic and heavy meson
effective theory (HMET) convention, respectively. They can be
related with each other by
\begin{equation}
|\mathcal{H}(p)\rangle=\sqrt{m_H}\left[|\mathcal{H}(v)\rangle+\mathcal{O}(1/m_H)\right].
\end{equation}
Then, in the framework of HMET, Eq.~\eqref{FormPV} can be
reexpressed as
\begin{equation}\label{FormPVNon}
\langle P(p^\prime)|J_{em}^\mu|V(p,\varepsilon_V)\rangle=e\sqrt{m_V
m_P} \mu^\prime(q^2) \epsilon^{\mu\nu\alpha\beta} v_\nu q_\alpha
\varepsilon_{V\beta},
\end{equation}
where the recoil effect is negligible in the above equation.

With the above preparation, one can easily get the expression of the
decay rate,
\begin{equation}%\label{FormPV}
\Gamma\left[V\to P\gamma\right]=\frac{1}{3}\int d\Omega_{\hat{q}}\frac{1}{32\pi^2}\frac{|\boldsymbol{q}|}{m_V^2}\sum|\mathcal{M}|^2,%\frac{\alpha}{3}\frac{m_P}{m_V}\left|\mathcal{F}(q^2)\right|^2|\mathbf{q}|^3,
\end{equation}
where $\mathcal{M}$ represents the transition amplitude, and a sum
over the final-state photon polarization and an average over the
initial $V$ polarization has been performed.

Explicitly, we have
\begin{equation}%\label{FormPV}
\Gamma\left[V\to
P\gamma\right]=\frac{\alpha}{3}\frac{m_P}{m_V}\left|\mu^\prime(0)\right|^2|\boldsymbol{q}|^3,
\end{equation}
where $\alpha=1/137$ is the fine-structure constant. The transition
magnetic moment $\mu_{V\to P\gamma}$ can be defined as
\begin{equation}%\label{FormPV}
\mu_{V\to P\gamma}=\frac{e}{2}\mu^\prime(0).
\end{equation}

In the following, we derive the magnetic moment of a vector meson.
The matrix element of $J_{em}^\mu(q^2)$ are defined in terms of the
standard Lorentz covariant decomposition~\cite{Arnold:1979cg},
\begin{eqnarray}%\label{FormPV}
\mathcal{G}^\mu(q^2)&=&\langle V(p^\prime,\varepsilon^{\prime\ast})|J_{em}^\mu(q^2)|V(p,\varepsilon)\rangle \nonumber\\
&=&-\mathcal{G}_1(q^2)(\varepsilon\cdot\varepsilon^{\prime\ast})(p+p^\prime)^\mu \nonumber\\
&&+\mathcal{G}_2(q^2)\left[(\varepsilon\cdot q)\varepsilon^{\prime\ast\mu}-(\varepsilon^{\prime\ast}\cdot q)\varepsilon^{\mu}\right]\nonumber\\
&&+\mathcal{G}_3(q^2)\frac{(\varepsilon\cdot
q)(\varepsilon^{\prime\ast}\cdot q)}{2m_V^2}(p+p^\prime)^\mu.
\end{eqnarray}
This expression can be simplified under Breit frame. In our
calculations, we define
\begin{eqnarray}%\label{FormPV}
&q^\mu=(p-p^\prime)^\mu=(0,\boldsymbol{Q}),\quad\boldsymbol{Q}=Q\hat{z},\quad p^\mu=(p^0, \frac{1}{2}\boldsymbol{Q}),\nonumber\\
&p^{\prime\mu}=(p^0, -\frac{1}{2}\boldsymbol{Q}),\quad
-q^2=Q^2\ge0,\quad p^0=\sqrt{m_V^2+\frac{1}{4}Q^2}.\nonumber
\end{eqnarray}
A straightforward derivation under Breit frame gives the time
component of $\mathcal{G}^\mu(q^2)$ as
\begin{eqnarray}\label{GTime}
\mathcal{G}^0(Q^2)&=&2p^0\Bigg\{\mathcal{G}_C(Q^2)(\boldsymbol{\varepsilon}\cdot\boldsymbol{\varepsilon}^{\prime\ast})+\frac{\mathcal{G}_Q(Q^2)}{2m_V^2}\Big[(\boldsymbol{\varepsilon}\cdot\boldsymbol{Q})(\boldsymbol{\varepsilon}^{\prime\ast}\cdot\boldsymbol{Q})\nonumber\\
&&-\frac{1}{3}(\boldsymbol{\varepsilon}\cdot\boldsymbol{\varepsilon}^{\prime\ast})Q^2\Big]\Bigg\},
\end{eqnarray}
where $\mathcal{G}_C$ and $\mathcal{G}_Q$ represent charge and
quadrupole form factors, respectively. In deriving
Eq.~\eqref{GTime}, we have used the transverse condition of the
initial and final state polarization vectors, i.e.,
$p\cdot\varepsilon=0$, and
$p^\prime\cdot\varepsilon^{\prime\ast}=0$.

Similarly, the space component of $\mathcal{G}^\mu(q^2)$ is
\begin{eqnarray}\label{GSpace}
\boldsymbol{\mathcal{G}}(Q^2)&=&\mathcal{G}_2(Q^2)\left[(\boldsymbol{\varepsilon}^{\prime\ast}\cdot\boldsymbol{Q})\boldsymbol{\varepsilon}-(\boldsymbol{\varepsilon}\cdot\boldsymbol{Q})\boldsymbol{\varepsilon}^{\prime\ast}\right]\nonumber\\
&=&2p^0\frac{\mathcal{G}_M(Q^2)}{2m_V}\left[(\boldsymbol{\varepsilon}^{\prime\ast}\cdot\boldsymbol{Q})\boldsymbol{\varepsilon}-(\boldsymbol{\varepsilon}\cdot\boldsymbol{Q})\boldsymbol{\varepsilon}^{\prime\ast}\right],
\end{eqnarray}
where $\mathcal{G}_M$ is the magnetic dipole form factor. The
expressions of $\mathcal{G}_C$, $\mathcal{G}_Q$ and $\mathcal{G}_M$
read
\begin{eqnarray}%\label{GSpace}
\mathcal{G}_C&=&\mathcal{G}_1+\frac{2}{3}\eta\mathcal{G}_Q,\nonumber\\
\mathcal{G}_Q&=&\mathcal{G}_3+\mathcal{G}_2(1+\eta)^{-1}+\frac{1}{2}\mathcal{G}_1(1+\eta)^{-1},\nonumber\\
\mathcal{G}_M&=&\mathcal{G}_2,
\end{eqnarray}
where $\eta={Q^2}/{(4m_V^2)}$.

\section{effective Lagrangians}\label{sec3}

\subsection{The leading order chiral Lagrangians}
We first introduce the Lagrangian of Goldstone bosons and photon.
The octet of the light pseudoscalar field is represented by the
field $U(x)=e^{i\phi/f_\phi}$ with
\begin{equation}
\phi=\left( \begin{array}{ccc}{\pi^{0}+\frac{1}{\sqrt{3}} \eta} &
{\sqrt{2} \pi^{+}} & {\sqrt{2} K^{+}} \\ {\sqrt{2} \pi^{-}} &
{-\pi^{0}+\frac{1}{\sqrt{3}} \eta} & {\sqrt{2} K^{0}} \\ {\sqrt{2}
K^{-}} & {\sqrt{2} \bar{K}^{0}} & {-\frac{2}{\sqrt{3}}
\eta}\end{array}\right),
\end{equation}
where the $\eta$ field denotes the octet $\eta_8$. In the SU(3) quark model, the $\eta$ meson is regarded as the mixing of the octet $\eta_8$ and singlet $\eta_0$ with $|\eta\rangle=\cos\theta|\eta_8\rangle-\sin\theta|\eta_0\rangle$ \cite{Wang:2016qmz}, where $\theta\simeq-19.1^\circ$ is determined by the experimental measurements \cite{Coffman:1988ve,Jousset:1988ni}. Because the mixing angle is not very large and the $\eta$ field only serves as the quantum fluctuations in the loops, so the mixing effect is ignored in our calculations.

The definitions of the chiral connection and axial-vector current
are
\begin{eqnarray}\label{GammaU}
\Gamma_{\mu} &\equiv& \frac{1}{2}\left[u^{\dagger}\left(\partial_{\mu}-i r_{\mu}\right) u+u\left(\partial_{\mu}-i l_{\mu}\right) u^{\dagger}\right], \\
u_{\mu} &\equiv& \frac{i}{2}\left[u^{\dagger}\left(\partial_{\mu}-i
r_{\mu}\right) u-u\left(\partial_{\mu}-i l_{\mu}\right)
u^{\dagger}\right],
\end{eqnarray}
where
\begin{eqnarray}
u^2=U=\exp\left(\frac{i\phi}{f_\phi}\right),\quad\quad
r_\mu=l_\mu=-eQA_\mu,
\end{eqnarray}
and $Q=Q_l=\mathrm{diag}(2/3,-1/3,-1/3)$ represents the electric
charge matrix of the light current $J^\ell_\mu$,
\begin{eqnarray}
J^\ell_\mu=\frac{2}{3}\bar{u}\gamma_\mu
u-\frac{1}{3}\bar{d}\gamma_\mu d-\frac{1}{3}\bar{s}\gamma_\mu s.
\end{eqnarray}
$f_\phi$ is the decay constant of the light pseudoscalar mesons. The
experimental value of $f_\phi$ for $\phi=\pi,K,\eta$ are
$f_\pi=92.4$ MeV, $f_K=113$ MeV, and $f_\eta=116$ MeV, respectively.

The leading order $[\mathcal{O}(p^2)]$ Lagrangian for the
interaction of the light pseudoscalars and photon
read~\cite{Li:2017cfz,Li:2017pxa,Meng:2017dni}
\begin{equation}
\mathcal{L}_{\phi\gamma}^{(2)}=\frac{f_{\phi}^{2}}{4}
\operatorname{Tr}\left[\nabla_{\mu} U\left(\nabla^{\mu}
U\right)^{\dagger}\right],
\end{equation}
where
\begin{equation}
\nabla_{\mu} U=\partial_{\mu} U-i r_{\mu} U+i U l_{\mu}.
\end{equation}
We use $\operatorname{Tr}(X)$ and $\langle X\rangle$ to denote the
trace for $X$ in flavor space and spinor space, respectively.

We construct the effective Lagrangian for the heavy mesons with the
superfield $\mathcal{H}$. For a heavy meson composed of a heavy
antiquark $\bar{Q}$ and a light quark $q$, the superfield
$\mathcal{H}$ is defined as
\begin{eqnarray}
\mathcal{H}&=&\left(P_{\mu}^{*} \gamma^{\mu}+i P \gamma_{5}\right) \frac{1-\slashed{v}}{2},\nonumber\\
\bar{\mathcal{H}}&=&\gamma^{0} \mathcal{H}^{\dagger}
\gamma^{0}=\frac{1-\slashed{v}}{2}\left(P_{\mu}^{* \dagger}
\gamma^{\mu}+i P^{\dagger} \gamma_{5}\right),
\end{eqnarray}
where for the charmed mesons
\begin{eqnarray}
P=(\bar{D}^0,D^-,D_s^-),~P^\ast=(\bar{D}^{0\ast},D^{\ast-},D_s^{\ast-}),
\end{eqnarray}
and for the bottom mesons
\begin{eqnarray}
P=(B^+,B^0,B_s^0),~P^\ast=(B^{\ast+},B^{\ast0},B_s^{\ast0}).
\end{eqnarray}

The leading order Lagrangians for describing the interactions between the heavy matter field and light pseudoscalars are~\cite{Wise:1992hn,Manohar:2000dt}
\begin{equation}\label{LHHphi}
\mathcal{L}_{H \phi}^{(1)}=-i\langle\bar{\mathcal{H}} v \cdot
\mathcal{D} \mathcal{H}\rangle-\frac{1}{8}
\Delta\langle\bar{\mathcal{H}} \sigma^{\mu \nu} \mathcal{H}
\sigma_{\mu \nu}\rangle+ g\langle\bar{\mathcal{H}} \slashed{u}
\gamma_{5} \mathcal{H}\rangle,
\end{equation}
where the covariant derivative $\mathcal{D}_\mu=\partial_\mu+\Gamma_\mu$. Here, the electric
charge matrix in the $\Gamma_\mu$ should be replaced by those
corresponding to the heavy mesons. For instance,
$Q=Q_D=\mathrm{diag}(0,-1,-1)$ for $(\bar{D}^{\ast0}, D^{\ast-},
D_s^{\ast-})$, and $Q=Q_B=\mathrm{diag}(1,0,0)$ for $(B^{\ast+},
B^{\ast0}, B_s^{\ast0})$, respectively. The second term in
Eq.~\eqref{LHHphi} is due to the mass difference between $P$ and
$P^\ast$, and $\Delta=m_{P^\ast}-m_P$ stands for the mass splitting.
$g$ represents the axial coupling constant. For the $D$ meson, its value can
be extracted by the partial decay width of $D^{\ast+}\to D^0\pi^+$ \cite{Tanabashi:2018oca,Ahmed:2001xc}, while for
the $B$ meson, $g$ can only be determined via the theoretical method, such as the quark model \cite{Casalbuoni:1996pg} and
lattice QCD \cite{Ohki:2008py,Detmold:2012ge}.

We also need the Lagrangians to describe the (transition) magnetic
moments at the tree level, which can be written as \cite{Li:2017cfz,Li:2017pxa,Meng:2017dni}
\begin{eqnarray}\label{LHHgamma2}
\mathcal{L}_{H\gamma}^{(2)}=\tilde{a}\langle\bar{\mathcal{H}}\sigma^{\mu\nu}\tilde{f}_{\mu\nu}^+\mathcal{H}\rangle+a\langle\mathcal{H}\sigma^{\mu\nu}\bar{\mathcal{H}}\rangle\operatorname{Tr}(f_{\mu\nu}^+),
\end{eqnarray}
where $\tilde{a}$ and $a$ are two LECs. The first and second terms
correspond to the contributions from the light quark and heavy
antiquark, respectively. The field strength tensor
$\tilde{f}_{\mu\nu}^+$ and $f_{\mu\nu}^+$ are defined as
\begin{eqnarray}\label{fmn}
f_{\mu\nu}^R&=&f_{\mu\nu}^L=-eQ\left(\partial_\mu A_\nu-\partial_\nu A_\mu\right),\nonumber\\
f_{\mu\nu}^{\pm}&=&u^\dag f_{\mu\nu}^R u \pm u f_{\mu\nu}^L u^\dag,\nonumber\\
\tilde{f}_{\mu\nu}^{\pm}&=&f_{\mu\nu}^{\pm}-\frac{1}{3}\mathrm{Tr}(f_{\mu\nu}^{\pm}),
\end{eqnarray}
where $Q=Q_D$ for the $D$ mesons and $Q=Q_B$ for the $B$ mesons,
respectively. From Eq.~\eqref{fmn} we can see that
$\tilde{f}_{\mu\nu}^{+}$ is proportional to $Q_l$ and traceless.
$f_{\mu\nu}^{+}$ is not traceless because it contains the electric
charge matrix of the heavy mesons. One can also understand
Eq.~\eqref{LHHgamma2} from the standpoint of group representation
theory. Recall that $3\otimes\bar{3}=1\oplus8$, and the operator
$\tilde{f}_{\mu\nu}^{+}$ transforms as the adjoint representation.
Thus the two terms in Eq.~\eqref{LHHgamma2} correspond to
$8\otimes8\to1$ and $1\otimes1\to1$, respectively.

In the following, we construct the Lagrangian for the interactions
of the heavy mesons and light pseudoscalar mesons at
$\mathcal{O}(p^2)$, which will contribute to the $\mathcal{O}(p^4)$
magnetic moment at the one-loop level \cite{Li:2017cfz,Li:2017pxa,Meng:2017dni},
\begin{eqnarray}\label{Hphiphi2}
\mathcal{L}_{H\phi\phi}^{(2)}&=&ib\langle\bar{\mathcal{H}}\sigma^{\mu\nu}[u_\mu,u_\nu]\mathcal{H}\rangle.
\end{eqnarray}
Actually, the tensor structure sandwiched between
$\bar{\mathcal{H}}$ and $\mathcal{H}$ in Eq.~\eqref{Hphiphi2} can
also be $\{u_\mu,u_\nu\}$ and $\mathrm{Tr}(u_\mu u_\nu)$. For the
$\mathrm{SU}(3)$ group representations,
\begin{eqnarray}
3\otimes\bar{3}&=&1\oplus8,\nonumber\\
8\otimes8&=&1\oplus8_1\oplus8_2\oplus10\oplus\overline{10}\oplus27.
\end{eqnarray}
The axial-vector current $u_\mu$ (or $u_\nu$) transforms as the
adjoint representation, thus $\mathrm{Tr}(u_\mu u_\nu)$,
$[u_\mu,u_\nu]$ and $\{u_\mu,u_\nu\}$ belong to $1$, $8_1$ and $8_2$
flavor representations, respectively. But $\mathrm{Tr}(u_\mu u_\nu)$
and $\{u_\mu,u_\nu\}$ would vanish when they are contracted with
$\sigma^{\mu\nu}$ because of the symmetric Lorentz indices $\mu$ and
$\nu$. Therefore, only one independent term containing
$[u_\mu,u_\nu]$ survives in Eq.~\eqref{Hphiphi2}.

\subsection{The next-to-leading order chiral Lagrangians}
The electromagnetic chiral Lagrangians at $\mathcal{O}(p^3)$ read \cite{Wang:2018cre}
\begin{eqnarray}\label{LHgamma3}
\mathcal{L}_{H\gamma}^{(3)}&=&-i\tilde{c}\langle\bar{\mathcal{H}}\sigma^{\mu\nu}v\cdot\nabla\tilde{f}_{\mu\nu}^+\mathcal{H}\rangle-ic\langle\mathcal{H}\sigma^{\mu\nu}\bar{\mathcal{H}}\rangle v\cdot\nabla\mathrm{Tr}(f_{\mu\nu}^+).\nonumber\\
\end{eqnarray}
The structure is similar to those in Eq.~\eqref{LHHgamma2}. The
possible contributions that include covariant derivative
$\mathcal{D}_\mu$ can be absorbed into the LECs $\tilde{c}$ and $c$
with the equation of motion of the heavy mesons. Meanwhile, the
contributions from Eq.~\eqref{LHgamma3} can be absorbed into
Eq.~\eqref{LHHgamma2} by renormalizing the LECs $\tilde{a}$ and $a$,
i.e.,
\begin{eqnarray}
\tilde{a}\rightarrowtail\tilde{a}+\tilde{c}v\cdot q,\quad\quad
a\rightarrowtail a+c v\cdot q.
\end{eqnarray}

\subsection{The next-to-next-to-leading order chiral Lagrangians}
At this order, we also employ group representation methods to
construct the electromagnetic chiral Lagrangians (one can find the
possible flavor structures in Table~\ref{Flavor_Structure}), the
detailed form reads \cite{Li:2017cfz,Li:2017pxa,Meng:2017dni}
\begin{eqnarray}\label{Hgamma4}
\mathcal{L}_{H\gamma}^{(4)}&=&\tilde{d}\langle\mathcal{H}\sigma^{\mu\nu}\tilde{\chi}_+\bar{\mathcal{H}}\rangle \mathrm{Tr}(f_{\mu\nu}^{+})+\bar{d}\langle\bar{\mathcal{H}}\sigma^{\mu\nu}\mathcal{H}\rangle\mathrm{Tr}(\tilde{f}_{\mu\nu}^{+}\tilde{\chi}_+)\nonumber\\
&&+d\langle\bar{\mathcal{H}}\sigma^{\mu\nu}\{\tilde{\chi}_+,\tilde{f}_{\mu\nu}^+\}\mathcal{H}\rangle,
\end{eqnarray}
where a spurion $\chi_{\pm}$ is introduced as
\begin{eqnarray}
\chi&=&2B_0\mathrm{diag}(m_u,m_d,m_s)=\mathrm{diag}(m_\pi^2,m_\pi^2,2m_K^2-m_\pi^2),\nonumber\\
\chi_{\pm}&=&u^\dag\chi u^\dag\pm u\chi^\dag u.\nonumber
\end{eqnarray}
At the leading order,
\begin{eqnarray}
\chi_+&=&\mathrm{diag}(2m_\pi^2,2m_\pi^2,4m_K^2-2m_\pi^2),\nonumber\\
\tilde{\chi}_+&=&\chi_+-\frac{1}{3}\Tr(\chi_+).
\end{eqnarray}
In principle, there should be six independent terms in
Eq.~\eqref{Hgamma4} as the possible flavor structures listed in
Table~\ref{Flavor_Structure}. However, the terms
$\mathrm{Tr}(\chi_+)\mathrm{Tr}(f_{\mu\nu}^+)$ and
$\mathrm{Tr}(\chi_+) \tilde{f}_{\mu\nu}^+$ can also be absorbed into
Eq.~\eqref{LHHgamma2} by renormalizing $\tilde{a}$ and $a$,
respectively. Another term $[\tilde{\chi}_+,\tilde{f}_{\mu\nu}^+]$
vanishes since both $\tilde{\chi}_+$ and $\tilde{f}_{\mu\nu}^+$ are
diagonal matrices at the leading order. Therefore, only three terms
are retained in Eq.~\eqref{Hgamma4}.
\begin{table*}[htbp]
\renewcommand{\arraystretch}{1.7}
 \tabcolsep=1.8pt
\caption{The possible flavor structures of the $\mathcal{O}(p^4)$
Lagrangians that contribute to the magnetic
moments.}\label{Flavor_Structure} \setlength{\tabcolsep}{4.6mm} {
\begin{tabular}{c|cccccc}
\hline\hline
Group representations& $1\otimes1\to1$ &$1\otimes8\to8$ &$8\otimes1\to8$ &$8\otimes8\to1$ &$8\otimes8\to8_1$ &$8\otimes8\to8_2$\\
\hline
Flavor structures& $\mathrm{Tr}(\chi_+)\mathrm{Tr}(f_{\mu\nu}^+)$ &$\mathrm{Tr}(\chi_+) \tilde{f}_{\mu\nu}^+$ &$\tilde{\chi}_+\mathrm{Tr}(f_{\mu\nu}^+)$ &$\mathrm{Tr}(\tilde{\chi}_+\tilde{f}_{\mu\nu}^+)$ &$[\tilde{\chi}_+,\tilde{f}_{\mu\nu}^+]$ &$\{\tilde{\chi}_+,\tilde{f}_{\mu\nu}^+\}$\\
\hline\hline
\end{tabular}
}
\end{table*}

\section{Radiative transitions}\label{sec4}

\subsection{Power counting and analytical expressions for the transition from factors}
The standard power counting scheme gives the chiral order of a
Feynman diagram as
\begin{eqnarray}
\mathcal{O}=4N_L-2I_M-I_H+\sum_n nN_n,
\end{eqnarray}
where $N_L$, $I_M$ and $I_H$ are the numbers of loops, internal
light pseudoscalar lines and internal heavy meson lines,
respectively. $N_n$ is the number of vertices governed by the $n$th
order Lagrangians. Usually, the order of the (transition) magnetic
moment is
\begin{eqnarray}
\mathcal{O}_\mu=\mathcal{O}-1.
\end{eqnarray}
Therefore, the transition form factors of $V\to P\gamma$ can be
expressed as follows,
\begin{eqnarray}\label{muVPgamma}
\mu_{V\to
P\gamma}^\prime=\left[\mu^{\prime(1)}_{\mathrm{tree}}\right]+\left[\mu^{\prime(2)}_{\mathrm{loop}}\right]+\left[\mu^{\prime(3)}_{\mathrm{tree}}+\mu^{\prime(3)}_{\mathrm{loop}}\right],
\end{eqnarray}
where the numbers in the parentheses are the chiral order
$\mathcal{O}_\mu$.
\begin{figure}[hptb]
%\vspace{0.3cm}
    %\setlength{\abovecaptionskip}{0.1cm}
    %\setlength{\belowcaptionskip}{-0.4cm}
    \scalebox{1.0}{\includegraphics[width=\columnwidth]{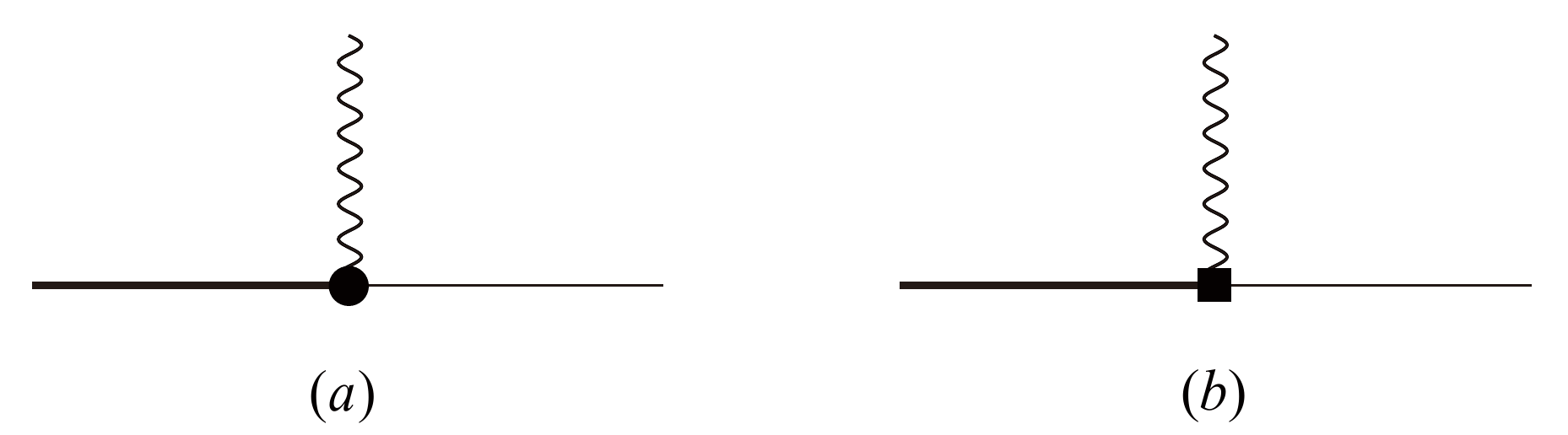}}
    \caption{The diagrams for the $V\to P\gamma$ transitions at the tree level. The thick solid, thin solid, and wiggly lines represent the vector meson $V$, pseudoscalar meson $P$, and photon $\gamma$, respectively. The solid circle and solid square in figures $(a)$ and $(b)$ correspond to the $\mathcal{O}(p^2)$ and $\mathcal{O}(p^4)$ vertices, respectively.\label{Tree_Diagrams}}
\end{figure}

We first study the $V\to P\gamma$ transitions. The tree diagrams are
illustrated in Fig.~\ref{Tree_Diagrams}. Expanding the Lagrangians
in Eqs.~\eqref{LHHgamma2} and~\eqref{Hgamma4} we can easily get the
transition amplitudes of Figs.~\ref{Tree_Diagrams}$(a)$
and~\ref{Tree_Diagrams}$(b)$, respectively. We can extract the
$q^2$-independent form factor $\mu^\prime$ at the tree level by
comparing the transition amplitudes with Eqs.~\eqref{FormPV}
and~\eqref{FormPVNon}. The expressions read
\begin{eqnarray}
\mu^{\prime(a)}_{\bar{D}^{\ast0}\to \bar{D}^0\gamma}&=&\frac{16}{3}(\tilde{a}-3a),\label{mupDast0L}\\
\mu^{\prime(a)}_{D^{\ast-}\to D^-\gamma}&=&-\frac{8}{3}(\tilde{a}+6a),\\
\mu^{\prime(a)}_{D_s^{\ast-}\to D_s^-\gamma}&=&-\frac{8}{3}(\tilde{a}+6a),\label{mupDsastmL}\\
\mu^{\prime(b)}_{\bar{D}^{\ast0}\to \bar{D}^0\gamma}&=&-\frac{32}{9}(m_K^2-m_\pi^2)(-6\tilde{d}+3\bar{d}+4d),\\
\mu^{\prime(b)}_{D^{\ast-}\to D^-\gamma}&=&-\frac{32}{9}(m_K^2-m_\pi^2)(-6\tilde{d}+3\bar{d}-2d),\\
\mu^{\prime(b)}_{D_s^{\ast-}\to
D_s^-\gamma}&=&-\frac{32}{9}(m_K^2-m_\pi^2)(12\tilde{d}+3\bar{d}+4d).
\end{eqnarray}
We show the analytical expressions for the $D$ mesons, and display
the expressions for the $B$ mesons in Appendix~\ref{appdixA}.

\begin{figure*}[hptb]
%\vspace{0.3cm}
    %\setlength{\abovecaptionskip}{0.1cm}
    %\setlength{\belowcaptionskip}{-0.4cm}
    \scalebox{1.9}{\includegraphics[width=\columnwidth]{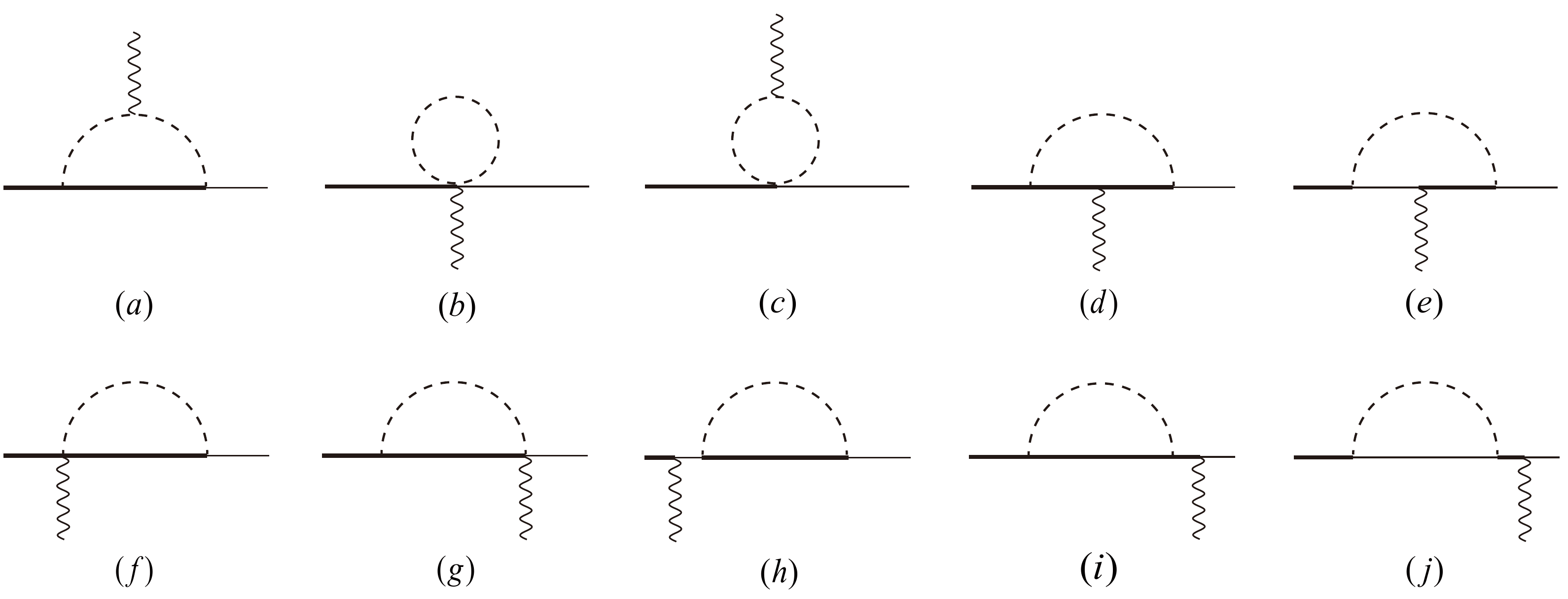}}
    \caption{The diagrams for the $V\to P\gamma$ transitions at the one-loop level, where the dashed line represents the light pseudoscalar mesons. Other notations are same as those in Fig.~\ref{Tree_Diagrams}.\label{Loop_Diagrams}}
\end{figure*}

The one-loop Feynman diagrams that contribute to the transition
processes are shown in Fig.~\ref{Loop_Diagrams}. Here, we need to
deal with the loop integrals when extracting the $q^2$-dependent
form factors from the transition amplitudes. Various types of loop
integrals $\mathcal{J}$ have been defined and given in
Appendix~\ref{appdixB}. In the following, we list the transition
form factors of
Figs.~\ref{Loop_Diagrams}$(a)$-\ref{Loop_Diagrams}$(j)$ with a
compact form, correspondingly.
\begin{eqnarray}
\mu^{\prime(a)}&=&\sum_\phi \mathcal{C}_\phi^{(a)}\frac{g^2}{f_\phi^2}\Big\{\mathcal{J}_{21}^T(m_\phi,\mathcal{E},q)\Big\}_r,\label{CphiAD}\\
\mu^{\prime(b)}&=&\sum_\phi \mathcal{C}_\phi^{(b)}\frac{\tilde{a}}{f_\phi^2}\Big\{\mathcal{J}_{0}^c(m_\phi)\Big\}_r,\\
\mu^{\prime(c)}&=&\sum_\phi \mathcal{C}_\phi^{(c)}\frac{b}{f_\phi^2}\Big\{\mathcal{J}_{22}^F(m_\phi,q)\Big\}_r,\\
\mu^{\prime(d)}&=&\sum_\phi \mathcal{C}_\phi^{(d)}\frac{g^2}{f_\phi^2}\Big\{\mathcal{J}_{22}^g(m_\phi,\mathcal{E},\mathcal{E}-q_0)\Big\}_r,\label{CphiDD}\\
\mu^{\prime(e)}&=&\sum_\phi \mathcal{C}_\phi^{(e)}\frac{g^2}{f_\phi^2}\Big\{\mathcal{J}_{22}^g(m_\phi,\mathcal{E}+\Delta,\mathcal{E}-q_0)\Big\}_r,\label{CphiED}\\
\mu^{\prime(f)}&=&\mu^{(g)}=0,\\
\mu^{\prime(h)}&=&\sum_\phi \mathcal{C}_\phi^{(h)}\frac{g^2}{f_\phi^2}\Big\{(1-D)\partial_\omega\mathcal{J}_{22}^a(m_\phi,\omega)\big|_{\omega\to-\Delta}\Big\}_r,\\
\mu^{\prime(i)+(j)}&=&\sum_\phi \mathcal{C}_\phi^{(ij)}\frac{g^2}{f_\phi^2}\Bigg\{\left[\partial_\omega\mathcal{J}_{22}^a(m_\phi,\omega)+2\partial_\delta\mathcal{J}_{22}^a(m_\phi,\delta)\right]\nonumber\\
&&\quad\quad\quad\quad\quad~\bigg|_{\omega\to\mathcal{E}+\Delta}^{\delta\to\mathcal{E}}\Bigg\}_r,\label{CphiIJD}
\end{eqnarray}
where the summations over $\phi$ denote the possible contributions
from the light pseudoscalars ($\phi$ could be $\pi$, $K$, $\eta$) in
the loops. $\mathcal{C}_\phi^{(x)}~(x=a,\dots,j)$ are the
flavor-dependent coefficients, and their values are given in
Tables~\ref{Flavor_Coe_D1}-\ref{Flavor_Coe_D2}. In the $\mathcal{J}$
functions, $m_\phi$ is the mass of the corresponding particle in the
loop. $\mathcal{E}$ is the residual energy of heavy mesons, which is
defined as $\mathcal{E}=E_{D^{(\ast)}}-m_{D^{(\ast)}}$.
$\mathcal{E}$ is set to be zero in our calculations. $q$ denotes the
transferred momentum carried by the photon. $D$ is the dimension in
dimensional regularization. $\{X\}_r$ represents the finite part of
$X$, which is defined in Appendix \ref{appdixB}. The coefficients
$\mathcal{C}_\phi^{(ij)}$ can be obtained via the relation
\begin{eqnarray}
\mathcal{C}_\phi^{(ij)}=-\mathcal{C}_\phi^{(h)}.
\end{eqnarray}
\begin{table*}[htbp]
\renewcommand{\arraystretch}{1.7}
 \tabcolsep=1.8pt
\caption{The flavor-dependent coefficients
$\mathcal{C}_\phi^{(x)}~(x=a,\dots,d)$ in
Eqs.~\eqref{CphiAD}-\eqref{CphiDD} for the $\bar{D}^\ast$
mesons.}\label{Flavor_Coe_D1} \setlength{\tabcolsep}{4.7mm} {
\begin{tabular}{c|ccccccccc}
\hline\hline
Decay modes& $\mathcal{C}_\pi^{(a)}$ &$\mathcal{C}_K^{(a)}$ &$\mathcal{C}_\pi^{(b)}$ &$\mathcal{C}_K^{(b)}$ &$\mathcal{C}_\pi^{(c)}$ &$\mathcal{C}_K^{(c)}$&$\mathcal{C}_\pi^{(d)}$&$\mathcal{C}_K^{(d)}$&$\mathcal{C}_\eta^{(d)}$\\
\hline
$\bar{D}^{\ast0}\to \bar{D}^0\gamma$&$2$&$2$&$-4$&$-4$&$4$&$4$&$24a$&$\frac{8}{3}(6a-\tilde{a})$&$\frac{8}{9}(3a+\tilde{a})$\\
$D^{\ast-}\to D^-\gamma$&$-2$&$0$&$4$&$0$&$-4$&$0$&$4(6a+\tilde{a})$&$\frac{8}{3}(6a-\tilde{a})$&$\frac{4}{9}(6a-\tilde{a})$\\
$D_s^{\ast-}\to D_s^-\gamma$&$0$&$-2$&$0$&$4$&$0$&$-4$&$0$&$\frac{8}{3}(12a+\tilde{a})$&$\frac{16}{9}(6a-\tilde{a})$\\
\hline\hline
\end{tabular}
}
\end{table*}
\begin{table*}[htbp]
\renewcommand{\arraystretch}{1.7}
 \tabcolsep=1.2pt
\caption{The flavor-dependent coefficients
$\mathcal{C}_\phi^{(x)}~(x=e,\dots,j)$ in
Eqs.~\eqref{CphiED}-\eqref{CphiIJD} for the $\bar{D}^\ast$
mesons.}\label{Flavor_Coe_D2} \setlength{\tabcolsep}{5.8mm} {
\begin{tabular}{c|cccccc}
\hline\hline
Decay modes& $\mathcal{C}_\pi^{(e)}$ &$\mathcal{C}_K^{(e)}$ &$\mathcal{C}_\eta^{(e)}$ &$\mathcal{C}_\pi^{(h)}$ &$\mathcal{C}_K^{(h)}$ &$\mathcal{C}_\eta^{(h)}$\\
\hline
$\bar{D}^{\ast0}\to \bar{D}^0\gamma$&$12a$&$\frac{4}{3}(6a+\tilde{a})$&$\frac{4}{9}(3a-\tilde{a})$&$2(\tilde{a}-3a)$&$\frac{4}{3}(\tilde{a}-3a)$&$\frac{2}{9}(\tilde{a}-3a)$\\
$D^{\ast-}\to D^-\gamma$&$2(6a-\tilde{a})$&$\frac{4}{3}(6a+\tilde{a})$&$\frac{2}{9}(6a+\tilde{a})$&$-(\tilde{a}+6a)$&$-\frac{2}{3}(\tilde{a}+6a)$&$-\frac{1}{9}(\tilde{a}+6a)$\\
$D_s^{\ast-}\to D_s^-\gamma$&$0$&$\frac{4}{3}(12a-\tilde{a})$&$\frac{8}{9}(6a+\tilde{a})$&$0$&$-\frac{4}{3}(\tilde{a}+6a)$&$-\frac{4}{9}(\tilde{a}+6a)$\\
\hline\hline
\end{tabular}
}
\end{table*}

\subsection{Estimation of the leading order LECs}
In $\mu^{\prime(1)}_{\mathrm{tree}}$, there exist two
$\mathcal{O}(p^2)$ LECs $\tilde{a}$ and $a$ (see
Eq.~\eqref{LHHgamma2}). Another $\mathcal{O}(p^2)$ LEC $b$ (see
Eq.~\eqref{Hphiphi2}) resides in $\mu^{\prime(3)}_{\mathrm{loop}}$.
In the following, we estimate the values of $\tilde{a}$, $a$ and $b$
with the quark model and resonance saturation model, respectively.
It is hard to determine the other higher-order LECs ($\tilde{d}$,
$\bar{d}$, and $d$) in $\mu^{\prime(3)}_{\mathrm{tree}}$ for the
moment because of very limited experimental data. Therefore, we
consider the contributions from $\mu^{\prime(3)}_{\mathrm{tree}}$ as
errors of our numerical results.

We first demonstrate how to determine $\tilde{a}$ and $a$ from the
scenario of constituent quark model. In this model, the transition
matrix element of $V\to P\gamma$ in the rest frame of the initial
state can be written as~\cite{Cheng:1992xi}
\begin{eqnarray}\label{VLemP1}
\langle P| \mathcal{L}_{\mathrm{em}}|V\rangle=2\sqrt{m_V m_P}\langle
P|\sum_i\frac{e_i}{2m_i}\boldsymbol{\sigma}|V\rangle\cdot\mathbf{B},
\end{eqnarray}
where $e_i$ and $m_i$ are the electric charge and mass of $i$th
quark in the heavy meson, $\boldsymbol{\sigma}$ and $\mathbf{B}$ are
the Pauli matrix and magnetic field, respectively. For simplicity,
we choose the direction of the magnetic field $\mathbf{B}$ along the
$z$ axis. In order to work out Eq.~\eqref{VLemP1}, we need the
flavor-spin wave functions of $V$ and $P$, which read
\begin{eqnarray}
|V\rangle&=&\frac{1}{\sqrt{2}}|\bar{Q}\uparrow q\downarrow+\bar{Q}\downarrow q\uparrow\rangle,\label{FSV}\\
|P\rangle&=&\frac{1}{\sqrt{2}}|\bar{Q}\uparrow
q\downarrow-\bar{Q}\downarrow q\uparrow\rangle.\label{FSP}
\end{eqnarray}
Inserting Eqs.~\eqref{FSV} and~\eqref{FSP} into Eq.~\eqref{VLemP1},
one can obtain
\begin{eqnarray}\label{VLemP2}
\langle P| \mathcal{L}_{\mathrm{em}}|V\rangle=2\sqrt{m_V
m_P}(\mu_{\bar{Q}}-\mu_q),
\end{eqnarray}
where $\mu_i=e_i/(2m_i)$. Matching Eq.~\eqref{VLemP2} with the
leading order transition amplitudes (i.e., replacing the $\mu^\prime(q^2)$ in Eq. \eqref{FormPVNon} with the expressions in Eqs. \eqref{mupDast0L}-\eqref{mupDsastmL}, and making use of $B^k(\boldsymbol{q})=-i\epsilon^{ijk}q^iA^j(\boldsymbol{q})$), one can easily get
\begin{eqnarray}\label{aat}
\tilde{a}=-\frac{1}{8m_q},\quad\quad\quad a=\frac{1}{24m_{\bar{Q}}},
\end{eqnarray}
where $m_q$ and $m_{\bar{Q}}$ are the masses of light constituent
quark and heavy antiquark in heavy mesons (in Appendix \ref{appdixC} we also give an estimation of the light quark mass with vector meson dominance model), respectively.
\begin{figure}[hptb]
%\vspace{0.3cm}
    %\setlength{\abovecaptionskip}{0.1cm}
    %\setlength{\belowcaptionskip}{-0.4cm}
    \scalebox{1.0}{\includegraphics[width=\columnwidth]{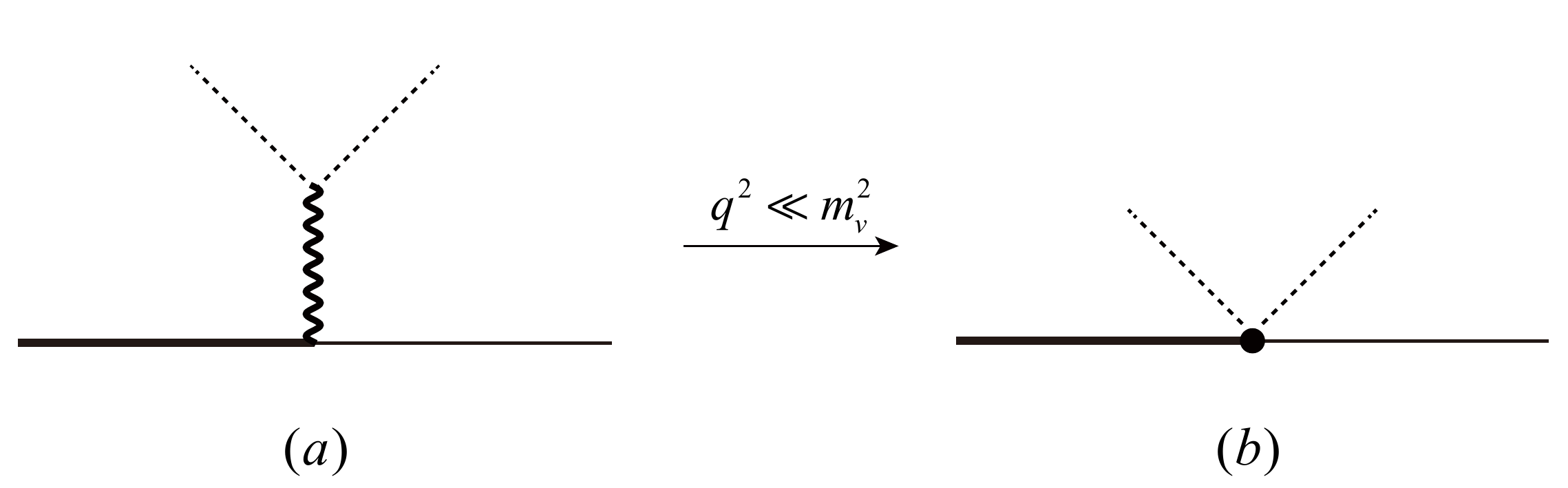}}
    \caption{A diagrammatic presentation of the resonance saturation scheme. The thick wiggly line in figure $(a)$ denotes the light vector meson $\rho$ or $\phi$, and other notations are same as those in Fig.~\ref{Loop_Diagrams}.\label{LEC_b}}
\end{figure}

Next, we evaluate the value of LEC $b$ in Eq.~\eqref{Hphiphi2} using
the resonance saturation model~\cite{Epelbaum:2001fm,Ecker:1988te}.
A diagrammatic presentation of the resonance saturation scheme is
illustrated in Fig.~\ref{LEC_b}. We need the interaction Lagrangians
for $VP\rho$ and $\rho\pi\pi~(\phi KK)$. The $VP\rho$ Lagrangian can
be obtained with local hidden symmetry~\cite{Casalbuoni:1996pg},
which reads
\begin{eqnarray}\label{LHrho}
\mathcal{L}_{H\rho}=i\beta\langle\bar{\mathcal{H}}v^\mu(\mathcal{V}_\mu-\rho_\mu)\mathcal{H}\rangle+i\lambda\langle\bar{\mathcal{H}}\sigma^{\mu\nu}F_{\mu\nu}(\rho)\mathcal{H}\rangle,
\end{eqnarray}
where
\begin{eqnarray}
F_{\mu\nu}(\rho)=\partial_\mu\rho_\nu-\partial_\nu\rho_\mu+[\rho_\mu,\rho_\nu],\quad\rho_\mu=i\frac{g_v}{\sqrt{2}}\hat{\rho}_\mu,
\end{eqnarray}
and
\begin{eqnarray}
\hat{\rho}^\mu=\left( \begin{array}{ccc}
\frac{\rho_0+\omega}{\sqrt{2}}&\rho^+&K^{\ast+} \\
\rho^-&\frac{-\rho_0+\omega}{\sqrt{2}}&K^{\ast0} \\
K^{\ast-}&\bar{K}^{\ast0}&\phi
\end{array} \right)^\mu.
\end{eqnarray}
The $\rho\pi\pi~(\phi KK)$ Lagrangian reads~\cite{Casalbuoni:1996pg}
\begin{eqnarray}\label{Lrhopi}
\mathcal{L}_{v\phi}=f_\phi^2
a\mathrm{Tr}(\Gamma_\mu^{(0)}\rho^\mu+\rho^\mu\Gamma_\mu^{(0)}),\quad
a=2,
\end{eqnarray}
where the expression of $\Gamma_\mu^{(0)}$ can be extracted from the
chiral connection defined in Eq.~\eqref{GammaU} by omitting the
photon field.

With the above preparations, we use the amplitude of
Fig.~\ref{LEC_b}$(a)$ governed by Lagrangians in Eqs.~\eqref{LHrho}
and~\eqref{Lrhopi} to match the amplitude of Fig.~\ref{LEC_b}$(b)$
depicted by the Lagrangian in Eq.~\eqref{Hphiphi2}. We can get the
$b$ explicitly
\begin{eqnarray}\label{bvalue}
b=-\frac{2\lambda g_v^2 f_\phi^2}{m_v^2},
\end{eqnarray}
where $g_v=5.8$, $\lambda=0.56~\mathrm{GeV}^{-1}$ \cite{Li:2012ss}. $m_v$ is the mass
of the exchanged light vector meson, such as
$m_\rho=0.77~\mathrm{GeV}$, and $m_\phi=1.02~\mathrm{GeV}$ \cite{Tanabashi:2018oca}. The sign
of $b$ is determined with the quark model.

The numerical values of the parameters
are~\cite{Tanabashi:2018oca,Li:2017cfz,Li:2017pxa,Meng:2017dni,Ahmed:2001xc,Ohki:2008py}
\begin{eqnarray}\label{Parameters}
m_\pi&=&0.139~\mathrm{GeV},\quad m_K=0.494~\mathrm{GeV},\quad m_\eta=0.548~\mathrm{GeV},\nonumber\\
m_u&=&m_d=0.336~\mathrm{GeV},\quad m_s=0.54~\mathrm{GeV},\nonumber\\
m_c&=&1.66~\mathrm{GeV},\quad m_b=4.73~\mathrm{GeV},\nonumber\\
g &=& \left\{ \begin{array}{ll}
0.59\pm0.01\pm0.07 & \textrm{for $D^\ast D\pi$ coupling}\\
0.516\pm0.05\pm0.033 & \textrm{for $B^\ast B\pi$ coupling}
\end{array} \right.,\nonumber\\
\Delta &=& \left\{ \begin{array}{ll}
0.142~\mathrm{GeV} & \textrm{for $m_{D^{\ast0}}-m_{D^0}$}\\
0.045~\mathrm{GeV} & \textrm{for $m_{B^{\ast0}}-m_{B^0}$}
\end{array} \right..
\end{eqnarray}
Since the masses of the mesons have been precisely measured in experiments \cite{Tanabashi:2018oca}, so we don't quote their minor errors. The masses of the constituent quarks are adopted from previous works \cite{Li:2017cfz,Li:2017pxa,Meng:2017dni}. Generally, it's hard to give the errors of the masses of the constituent quarks, because these values used in different quark models vary a lot some times. In this work, we try to give a conservative estimation by setting the $10\%\times m_q$ as the parameter errors. The axial constant $g$ for $D^\ast D\pi$ coupling is extracted from the experimental result of the CLEO Collaboration \cite{Ahmed:2001xc}. The $B^\ast B\pi$ coupling is quoted from the unquenched lattice result \cite{Ohki:2008py}.

\subsection{Numerical results and discussions}
With the parameters listed above, we first show the transition
magnetic moments of $V\to P\gamma$ calculated under SU(2) and SU(3)
symmetries\footnote{Here, SU(2) and SU(3) symmetries only imply the
effective Lagrangians are constructed under these two symmetries. The
SU(3) breaking effect is included explicitly in our calculations.
For example, we use the $m_{u,d,s}$ and the physical masses of $\pi$, $K$ and $\eta$ in
Eq. \eqref{Parameters} as inputs.} in the upper half parts of Tables~\ref{SU2_case}
and~\ref{SU3_case}, correspondingly. In Tables~\ref{SU2_case}
and~\ref{SU3_case}, the transition magnetic moments $\mu_{V\to
P\gamma}$ are given order by order. As expected, the convergence of
the chiral expansion in the SU(2) case is better than that in SU(3).
Besides, we also calculate the $\mu_{V\to P\gamma}$ with the mass
splitting $\Delta$ in the propagators of the loop diagrams kept and
unkept. The influence of $\Delta$ in the charm sector is more
significant than that in the bottom sector because the mass
difference of the charmed mesons is larger than that of the bottom
mesons.

In the SU(2) case, the mass splitting $\Delta$ only appears in the
loop diagrams. The transition magnetic moments at
$\mathcal{O}_\mu(p^1)$ remain unchanged no matter we choose
$\Delta=0$ or $\Delta\neq0$. At $\mathcal{O}_\mu(p^2)$, the
correction from the finite mass splitting ($\Delta\neq0$) is about
$40\%$ and $20\%$ for $\mu_{\bar{D}^\ast\to \bar{D}\gamma}$ and
$\mu_{B^\ast\to B\gamma}$, respectively. Such a correction is also
significant at $\mathcal{O}_\mu(p^3)$. Similar behavior is observed
in the SU(3) case at each order. In Table \ref{Each_Diagrams}, we show
the contribution of each loop diagram to the transition magnetic moment
of $\bar{D}^{\ast0}\to\bar{D}\gamma$ in different cases. The contributions
of the diagrams \ref{Loop_Diagrams}$(f)$ and \ref{Loop_Diagrams}$(g)$ vanish
in the heavy quark limit. Except for the diagrams \ref{Loop_Diagrams}$(b)$ and \ref{Loop_Diagrams}$(c)$,
other diagrams that contain the heavy meson internal line are effected by the
mass splitting $\Delta$. For the charmed vector mesons, $\Delta>m_\pi$, so the
loop integrals with the nonanalytic structures $\log\frac{y^2+m_\pi^2-\Delta^2-i\varepsilon}{\lambda^2}$
and $\sqrt{m_\pi^2-\Delta^2-i\varepsilon}$ (see Appendix \ref{appdixB}) would largely
impact the numerical result. This is vividly reflected in Table \ref{Each_Diagrams}. However, for
the bottom vector mesons, $\Delta\simeq1/3m_\pi$, so the influence of $\Delta$
on the bottom sector is not so obvious.

\begin{table*}
\renewcommand{\arraystretch}{1.9}
\caption{The contribution of each loop diagram to the transition magnetic moment of $\bar{D}^{\ast0}\to\bar{D}\gamma$ in different cases (in units of $\mu_N$).}\label{Each_Diagrams}
\setlength{\tabcolsep}{4.1mm}
\begin{tabular}{c|c|ccccccccc}
\hline\hline
\multicolumn{2}{c|}{Cases}&$(a)$&$(b)$&$(c)$&$(d)$&$(e)$&$(f)$&$(g)$&$(h)$&$(i+j)$\\
\hline
\multirow{2}{*}{SU(2)}&$\Delta=0$&$0.21$&$-0.085$&$0.062$&$0.012$&$0.006$&$0$&$0$&$-0.045$&$-0.053$\\
&$\Delta\neq0$&$0.29$&$-0.085$&$0.062$&$-0.0016$&$0.0021$&0&0&$0.073$&$-0.0082$\\
\hline
\multirow{2}{*}{SU(3)}&$\Delta=0$&$0.71$&$-0.37$&$0.27$&$0.088$&$-0.00017$&0&0&$-0.13$&$-0.19$\\
&$\Delta\neq0$&$0.81$&$-0.37$&$0.27$&$0.033$&$-0.0038$&0&0&$0.13$&$-0.19$\\
\hline
\hline
\end{tabular}
\end{table*}

The corresponding decay widths evaluated in different cases are
illustrated in Table~\ref{Numerical_Results}. The errors in our
calculations can stem from many sources, such as quark masses,
hadron masses, coupling constants, higher order contributions and so on.
As shown in the RPP \cite{Tanabashi:2018oca}, the errors of the hadron masses
appeared in this work are very small, so we ignore their effects.
Meanwhile, the axial coupling constant extracted from the experiments and
lattice QCD are also very small. Furthermore, the convergence of chiral
expansion works very well in our calculations. Therefore, we consider two main error sources.
The first one is the contribution of
the $\mathcal{O}(p^4)$ Lagrangians (see Eq.~\eqref{Hgamma4}). Since
the LECs in Eq.~\eqref{Hgamma4} cannot be fixed at present, we adopt
the nonanalytic dominance approximation to give an estimation of the
$\mathcal{O}(p^4)$ tree diagram~\cite{Wang:2018atz}. The second one
is the uncertainty from the quark models. For example, the masses of
constituent quarks are different in various models (see Table \ref{QuarkMass}). We take this
uncertainty into account. The change of the quark masses would lead
to a $10\%$ variation of the leading order LECs.

\begin{table}
\renewcommand{\arraystretch}{1.5}
\caption{The masses of the constituent quarks adopted in different works (in units of GeV).}\label{QuarkMass}
\setlength{\tabcolsep}{3.3mm}
\begin{tabular}{c|ccccc}
\hline
\hline
&$m_u$&$m_d$&$m_s$&$m_c$&$m_b$\\
\hline
Kamal \cite{Kamal:1992uv}&$0.34$&$0.34$&$0.55$&$1.8$&$\cdots$\\
Ebert \cite{Ebert:2002xz}&$0.33$&$0.33$&$0.5$&$1.55$&$4.88$\\
Cheng \cite{Cheng:1992xi}&$0.338$&$0.322$&$0.51$&$1.6$&$5.0$\\
Eichten \cite{Eichten:1979ms}&$0.335$&$0.335$&$0.45$&$1.84$&$5.17$\\
\hline
\hline
\end{tabular}
\end{table}

From Table~\ref{Numerical_Results}, we see that the decay rate for
$D^{\ast-}\to D^-\gamma$ calculated in different scenarios all
agrees with the experimental data. The branching ratios for the
other decay channels cannot be obtained due to the absence of the
total widths of these states in experiments at present. We also
compare our results with other model predictions, such as
light-front quark model~\cite{Choi:2007se}, relativistic independent
quark model~\cite{Priyadarsini:2016tiu}, relativistic quark
model~\cite{Ebert:2002xz} and QCD sum rules~\cite{Zhu:1996qy}. The
results in these literatures are consistent with our calculations.
Furthermore, the results from the extended Bag
model~\cite{Simonis:2016pnh,Simonis:2018rld}, lattice QCD
simulations~\cite{Becirevic:2009xp} and extended Nambu-Jona-Lasinio
model~\cite{Deng:2013uca} are also compatible with ours.
\iffalse
It's also illustrative to compare our results with the previous calculations
in the same framework. In Refs. \cite{Cho:1992nt,Cheng:1992xi}, Cho {\it et al}
and Cheng {\it et al} calculated the decay widths of $D^\ast\to D\gamma$ and $B^\ast\to B\gamma$
at the tree level in the heavy hadron chiral theory, respectively. Cho {\it et al} \cite{Cho:1992nt} showed $\Gamma_{D^{\ast0}\to D^0\gamma}=8.8\pm1.71$ keV,
$\Gamma_{D^{\ast+}\to D^+\gamma}=8.3\pm8.1$ keV with $m_c=1.7$ GeV, and $\Gamma_{B^{\ast0}\to B^0\gamma}=0.127\pm0.203$ keV,
$\Gamma_{B^{\ast+}\to B^+\gamma}=0.66\pm0.93$ keV with $m_b=5.0$ GeV. Cheng {\it et al} \cite{Cheng:1992xi} obtained $\Gamma_{D^{\ast0}\to D^0\gamma}=23$ keV,
$\Gamma_{D^{\ast+}\to D^+\gamma}=6$ keV, $\Gamma_{D_s^{\ast+}\to D_s^+\gamma}=2.4$ keV, and $\Gamma_{B^{\ast+}\to B^+\gamma}=0.84$ keV,
$\Gamma_{B^{\ast0}\to B^0\gamma}=0.28$ keV with $m_u=0.338$ GeV, $m_d=0.322$ GeV, $m_s=0.51$ GeV, $m_c=1.6$ GeV and $m_b=5.0$ GeV.
\fi
\begin{table*}
\renewcommand{\arraystretch}{1.7}
\caption{The radiative decay widths for $V\to P\gamma$ (in units of
keV). Br$_{\mathrm{expt}}$ and $\Gamma_{\mathrm{expt}}$ denote the
branching ratio and decay width measured in experiments.
$\Gamma_{1,\dots,4}$ are the model
predictions.}\label{Numerical_Results} \setlength{\tabcolsep}{1.8mm}
\begin{tabular}{c|cc|cc|ccccc}
\hline\hline
\multirow{2}{*}{Decay modes}&\multicolumn{2}{c|}{SU(2)}&\multicolumn{2}{c|}{SU(3)}&\multicolumn{5}{c}{Experimental data and model predictions}\\
&$\Delta=0$&$\Delta\neq0$&$\Delta=0$&$\Delta\neq0$&Br$_{\mathrm{expt}}$$\big|\Gamma_{\mathrm{expt}}$~\cite{Tanabashi:2018oca}&$\Gamma_1$~\cite{Choi:2007se}&$\Gamma_2$~\cite{Priyadarsini:2016tiu}&$\Gamma_3$~\cite{Ebert:2002xz}&$\Gamma_4$~\cite{Zhu:1996qy}\\
\hline
$\bar{D}^{\ast0}\to\bar{D}^0\gamma$&$30.0^{+7.3}_{-6.6}$&$23.9^{+5.0}_{-6.3}$&$22.9^{+8.2}_{-7.0}$&$16.2^{+6.5}_{-6.0}$&$(38.1\pm2.9)\%\big|\cdots$&$20.0\pm0.3$&26.5&$11.5$&$12.9\pm2$\\
$D^{\ast-}\to D^-\gamma$&$1.0^{+0.9}_{-0.6}$&$0.5^{+0.5}_{-0.4}$&$1.8^{+1.3}_{-0.9}$&$0.73^{+0.7}_{-0.3}$&$(1.6\pm0.4)\%\big|1.33\pm0.33$&$0.9\pm0.02$&0.93&$1.04$&$0.23\pm0.1$\\
$D_s^{\ast-}\to D_s^-\gamma$&$\cdots$&$\cdots$&$0.15^{+0.5}_{-0.1}$&$0.32^{+0.3}_{-0.3}$&$(94.2\pm0.7)\%\big|\cdots$&$0.18\pm0.01$&$0.21$&$0.19$&$0.13\pm0.05$\\
\hline
$B^{\ast+}\to B^+\gamma$&$0.75^{+0.2}_{-0.2}$&$0.71^{+0.2}_{-0.2}$&$0.63^{+0.2}_{-0.2}$&$0.58^{+0.2}_{-0.2}$&$\cdots\big|\cdots$&$0.4\pm0.03$&0.58&$0.19$&$0.13\pm0.03$\\
$B^{\ast0}\to B^0\gamma$&$0.19^{+0.05}_{-0.05}$&$0.18^{+0.05}_{-0.05}$&$0.25^{+0.06}_{-0.06}$&$0.23^{+0.06}_{-0.06}$&$\cdots\big|\cdots$&$0.13\pm0.01$&0.18&$0.07$&$0.38\pm0.06$\\
$B_s^{\ast0}\to B_s^0\gamma$&$\cdots$&$\cdots$&$0.05^{+0.03}_{-0.03}$&$0.04^{+0.03}_{-0.03}$&$\cdots\big|\cdots$&$0.068\pm0.017$&$0.12$&$0.05$&$0.22\pm0.04$\\
\hline\hline
\end{tabular}
\end{table*}

Up to now, only the full width of $D^{\ast\pm}$ and the branching
ratio of $D^{\ast\pm}\to D^{\pm}\gamma$ are available in
RPP~\cite{Tanabashi:2018oca}. The life time of $\bar{D}^{\ast0}$ and
$D_s^{\ast-}$ has not been measured yet. The convergence of the
chiral expansion for transition magnetic moments calculated in SU(3)
case with $\Delta\neq0$ is very reasonable. Therefore, as a
byproduct, we use the following relation with our results in SU(3)
and $\Delta\neq0$ as inputs to estimate the full widths of these two
states,
\begin{equation}\label{BrBr}
\frac{\operatorname{Br}(D^{*\pm} \rightarrow D^{ \pm}
\gamma)_{\mathrm{expt}}}{\mathrm{Br}(\bar{D}^{* 0} \rightarrow
\bar{D}^{0} \gamma)_{\mathrm{expt}}}=\frac{\Gamma(D^{* \pm}
\rightarrow D^{ \pm} \gamma)}{\Gamma(\bar{D}^{* 0} \rightarrow
\bar{D}^{0} \gamma)} \frac{\Gamma_{\mathrm{tot}}(\bar{D}^{*
0})}{\Gamma_{\mathrm{tot}}(D^{* \pm})},
\end{equation}
where the total width $\Gamma_{\mathrm{tot}}(\bar{D}^{\ast0})$ in
the above equation can be extracted with the predicted
$\Gamma(\bar{D}^{* 0} \rightarrow \bar{D}^{0} \gamma)$. Analogously,
$\Gamma_{\mathrm{tot}}(D_s^{\ast\pm})$ can also be calculated with
the same way as in the case of $\bar{D}^{\ast0}$. The full widths of
$\bar{D}^{\ast0}$ and $D_s^{\ast-}$ are estimated to be
\begin{eqnarray}\label{FullWidth}
\Gamma_{\mathrm{tot}}(\bar{D}^{\ast0})\simeq77.7^{+26.7}_{-20.5}~\mathrm{keV},\quad
\Gamma_{\mathrm{tot}}(D_s^{\ast-})\simeq0.62^{+0.45}_{-0.50}~\mathrm{keV},
\end{eqnarray}
respectively.

\section{Magnetic moments}\label{Mm}

The anomalous magnetic moments of nucleons reveal that the proton
and neutron are not elementary particles and they have internal
substructures. As in the case of nucleons, the magnetic moments of
$D^{\ast}$ and $B^{\ast}$ also encode important information of their
underlying substructures.

\subsection{Analytical expressions for the magnetic moments}
\begin{figure}[hptb]
%\vspace{0.3cm}
    %\setlength{\abovecaptionskip}{0.1cm}
    %\setlength{\belowcaptionskip}{-0.4cm}
    \scalebox{1.0}{\includegraphics[width=\columnwidth]{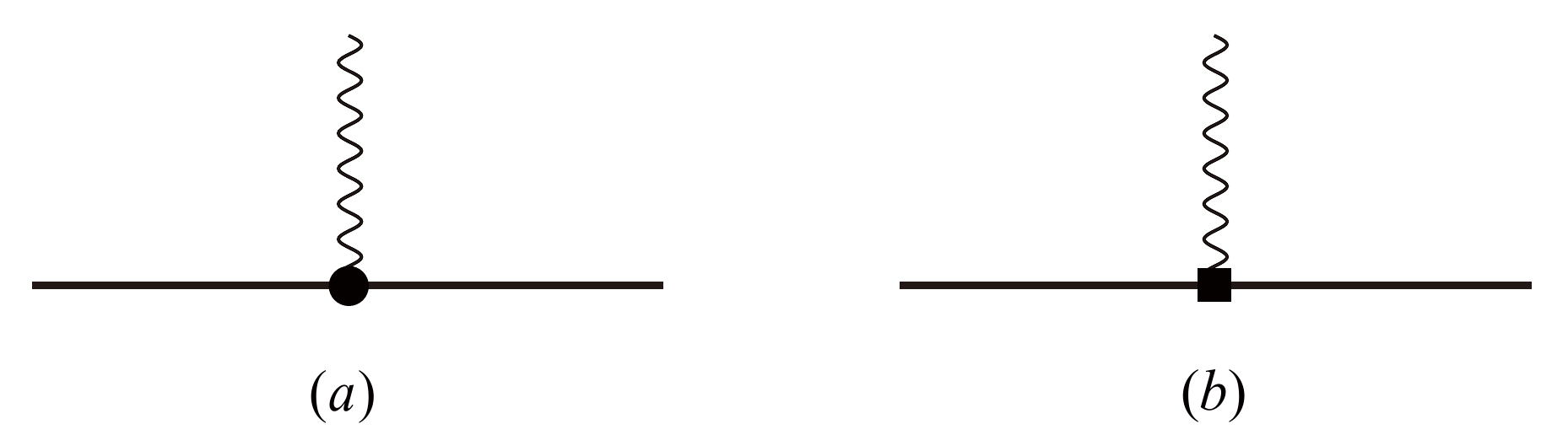}}
    \caption{Tree level Feynman diagrams that contribute to the magnetic moments of the heavy vector mesons. Notations are same as those in Fig.~\ref{Tree_Diagrams}. \label{Tree_Diagrams_Moments}}
\end{figure}

We have studied the radiative transitions $V\to P\gamma$ in previous
section. The decay rate for $D^{\ast-}\to D^-\gamma$ is consistent
with the experimental data. So we adopt the same set of parameters
to calculate the magnetic moments of the $D^{\ast}$ and $B^{\ast}$
mesons. The $\mathcal{O}(p^2)$ and $\mathcal{O}(p^4)$ tree level
Feynman diagrams that contribute to the magnetic moments are
displayed in Fig.~\ref{Tree_Diagrams_Moments}.

In the following, we write out the magnetic moments of the $D^\ast$
mesons from Figs.~\ref{Tree_Diagrams_Moments}$(a)$
and~\ref{Tree_Diagrams_Moments}$(b)$,
\begin{eqnarray}
\mu_{\bar{D}^{\ast0}}^{(a)}&=&-\frac{8}{3}e(\tilde{a}+3a),\label{muDastb0A}\\
\mu_{D^{\ast-}}^{(a)}&=&-\frac{4}{3}e(-\tilde{a}+6a),\label{muDastmA}\\
\mu_{D_s^{\ast-}}^{(a)}&=&-\frac{4}{3}e(-\tilde{a}+6a),\label{muDsastmA}\\
\mu_{\bar{D}^{\ast0}}^{(b)}&=&\frac{16}{9}e(m_K^2-m_\pi^2)(6\tilde{d}+3\bar{d}+4d),\\
\mu_{D^{\ast-}}^{(b)}&=&\frac{16}{9}e(m_K^2-m_\pi^2)(6\tilde{d}+3\bar{d}-2d),\\
\mu_{D_s^{\ast-}}^{(b)}&=&\frac{16}{9}e(m_K^2-m_\pi^2)(-12\tilde{d}+3\bar{d}+4d).
\end{eqnarray}

\begin{figure*}[hptb]
%\vspace{0.3cm}
    %\setlength{\abovecaptionskip}{0.1cm}
    %\setlength{\belowcaptionskip}{-0.4cm}
    \scalebox{1.9}{\includegraphics[width=\columnwidth]{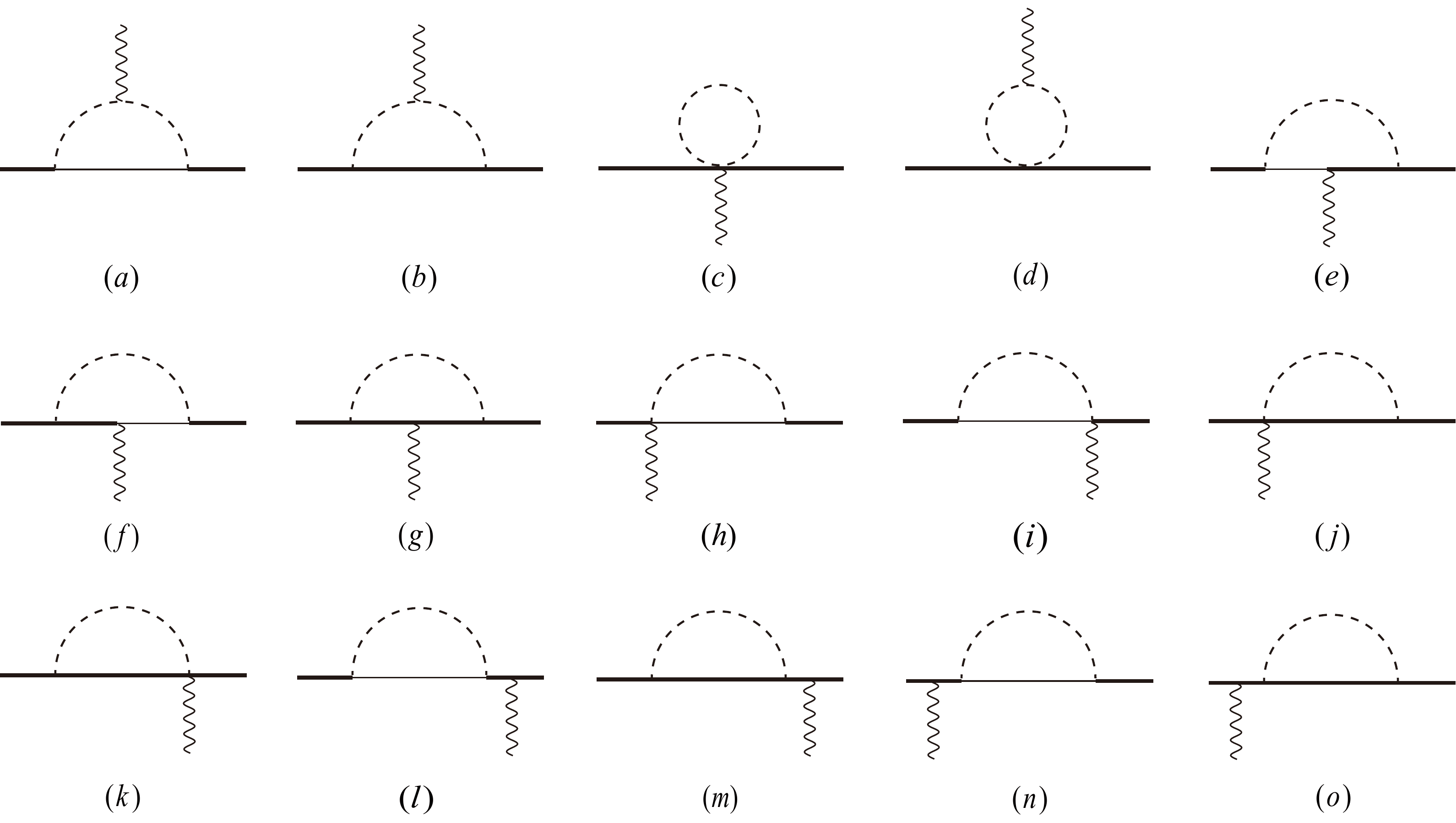}}
    \caption{One-loop Feynman diagrams that contribute to the magnetic moments of the heavy vector mesons. Notations are the same as those in Fig.~\ref{Loop_Diagrams}.\label{Loop_Diagrams_Moments}}
\end{figure*}

The magnetic moments from the one-loop diagrams in
Fig.~\ref{Loop_Diagrams_Moments} are given as
\begin{eqnarray}
\mu^{(a)}&=&\sum_\phi e\mathcal{C}_\phi^{(a)}\frac{g^2}{f_\phi^2}\mathcal{J}_{21}^T(m_\phi,\mathcal{E}+\Delta,q),\label{CphiADast}\\
\mu^{(b)}&=&\sum_\phi e\mathcal{C}_\phi^{(b)}\frac{g^2}{f_\phi^2}\mathcal{J}_{21}^T(m_\phi,\mathcal{E},q),\label{CphiBDast}\\
\mu^{(c)}&=&\sum_\phi e\mathcal{C}_\phi^{(c)}\frac{\tilde{a}}{f_\phi^2}\mathcal{J}_{0}^c(m_\phi),\\
\mu^{(d)}&=&\sum_\phi e\mathcal{C}_\phi^{(d)}\frac{b}{f_\phi^2}\mathcal{J}_{22}^F(m_\phi,q),\\
\mu^{(e)}&=&\sum_\phi e\mathcal{C}_\phi^{(e)}\frac{g^2}{f_\phi^2}\mathcal{J}_{22}^g(m_\phi,\mathcal{E}+\Delta,\mathcal{E}-q_0),\label{CphiEDast}\\
\mu^{(f)}&=&\sum_\phi e\mathcal{C}_\phi^{(f)}\frac{g^2}{f_\phi^2}\mathcal{J}_{22}^g(m_\phi,\mathcal{E},\mathcal{E}+\Delta-q_0),\\
\mu^{(g)}&=&\sum_\phi e\mathcal{C}_\phi^{(g)}\frac{g^2}{f_\phi^2}\mathcal{J}_{22}^g(m_\phi,\mathcal{E},\mathcal{E}-q_0),\label{CphiGDast}\\
\mu^{(h)}&=&\mu^{(i)}=\mu^{(j)}=\mu^{(k)}=0,\\
\mu^{(l)+(m)}&=&\mu^{(n)+(o)}=\sum_\phi e \mathcal{C}_\phi^{(lm)}\frac{g^2}{f_\phi^2}\Bigg\{\Big[\partial_\omega\mathcal{J}_{22}^a(m_\phi,\omega)\nonumber\\
&&+2\partial_\delta\mathcal{J}_{22}^a(m_\phi,\delta)\Big]\bigg|_{\omega\to\mathcal{E}+\Delta}^{\delta\to\mathcal{E}}\Bigg\}_r,\label{CphilmDast}
\end{eqnarray}
where the values of the coefficients
$\mathcal{C}_\phi^{(x)}~(x=a,\dots,o)$ for the $D^\ast$ mesons are
listed in Tables~\ref{Flavor_Coe_Dast1}-\ref{Flavor_Coe_Dast2}. In
Eqs.~\eqref{CphiADast} and~\eqref{CphiBDast}, we have used the
relation $\mathcal{J}_{31}^T=-\frac{1}{2}\mathcal{J}_{21}^T$ when
$q^2=0$ \cite{Wang:2018atz}. The unlisted coefficients
$\mathcal{C}_\phi^{(b)}$ and $\mathcal{C}_\phi^{(f)}$ can be
obtained by the relation
\begin{eqnarray}
\mathcal{C}_\phi^{(b)}=\mathcal{C}_\phi^{(a)},\quad\quad\quad
\mathcal{C}_\phi^{(f)}=\mathcal{C}_\phi^{(e)}.
\end{eqnarray}

\begin{table*}[htbp]
\renewcommand{\arraystretch}{1.7}
 \tabcolsep=1.8pt
\caption{The flavor-dependent coefficients
$\mathcal{C}_\phi^{(x)}~(x=a,c,d,e)$ in
Eqs.~\eqref{CphiADast}-\eqref{CphiEDast} for the $\bar{D}^\ast$
mesons.}\label{Flavor_Coe_Dast1} \setlength{\tabcolsep}{5.2mm} {
\begin{tabular}{c|ccccccccc}
\hline\hline
States& $\mathcal{C}_\pi^{(a)}$ &$\mathcal{C}_K^{(a)}$ &$\mathcal{C}_\pi^{(c)}$ &$\mathcal{C}_K^{(c)}$ &$\mathcal{C}_\pi^{(d)}$ &$\mathcal{C}_K^{(d)}$&$\mathcal{C}_\pi^{(e)}$&$\mathcal{C}_K^{(e)}$&$\mathcal{C}_\eta^{(e)}$\\
\hline
$\bar{D}^{\ast0}$&$-\frac{1}{2}$&$-\frac{1}{2}$&$2$&$2$&$-2$&$-2$&$6a$&$\frac{2}{3}(6a+\tilde{a})$&$\frac{2}{9}(3a-\tilde{a})$\\
$D^{\ast-}$&$\frac{1}{2}$&$0$&$-2$&$0$&$2$&$0$&$6a-\tilde{a}$&$\frac{2}{3}(6a+\tilde{a})$&$\frac{1}{9}(6a+\tilde{a})$\\
$D_s^{\ast-}$&$0$&$\frac{1}{2}$&$0$&$-2$&$0$&$2$&$0$&$\frac{2}{3}(12a-\tilde{a})$&$\frac{4}{9}(6a+\tilde{a})$\\
\hline\hline
\end{tabular}
}
\end{table*}
\begin{table*}[htbp]
\renewcommand{\arraystretch}{1.7}
 \tabcolsep=1.2pt
\caption{The flavor-dependent coefficients
$\mathcal{C}_\phi^{(x)}~(x=g,l+m)$ in
Eqs.~\eqref{CphiGDast}-\eqref{CphilmDast} for the $\bar{D}^\ast$
mesons.}\label{Flavor_Coe_Dast2} \setlength{\tabcolsep}{6.7mm} {
\begin{tabular}{c|cccccc}
\hline\hline
States& $\mathcal{C}_\pi^{(g)}$ &$\mathcal{C}_K^{(g)}$ &$\mathcal{C}_\eta^{(g)}$ &$\mathcal{C}_\pi^{(lm)}$ &$\mathcal{C}_K^{(lm)}$ &$\mathcal{C}_\eta^{(lm)}$\\
\hline
$\bar{D}^{\ast0}$&$6a$&$\frac{2}{3}(6a-\tilde{a})$&$\frac{2}{9}(3a+\tilde{a})$&$(\tilde{a}+3a)$&$\frac{2}{3}(\tilde{a}+3a)$&$\frac{1}{9}(\tilde{a}+3a)$\\
$D^{\ast-}$&$6a+\tilde{a}$&$\frac{2}{3}(6a-\tilde{a})$&$\frac{1}{9}(6a-\tilde{a})$&$\frac{1}{2}(6a-\tilde{a})$&$\frac{1}{3}(6a-\tilde{a})$&$\frac{1}{18}(6a-\tilde{a})$\\
$D_s^{\ast-}$&$0$&$\frac{2}{3}(12a+\tilde{a})$&$\frac{4}{9}(6a-\tilde{a})$&$0$&$\frac{2}{3}(6a-\tilde{a})$&$\frac{2}{9}(6a-\tilde{a})$\\
\hline\hline
\end{tabular}
}
\end{table*}

Analogous to the transition form factors $\mu_{V\to P\gamma}^\prime$
in Eq.~\eqref{muVPgamma}, the magnetic moments $\mu_V$ can be
written as
\begin{eqnarray}\label{muV}
\mu_{V}=\left[\mu^{(1)}_{\mathrm{tree}}\right]+\left[\mu^{(2)}_{\mathrm{loop}}\right]+\left[\mu^{(3)}_{\mathrm{tree}}+\mu^{(3)}_{\mathrm{loop}}\right],
\end{eqnarray}
where $\mu^{(1)}_{\mathrm{tree}}$, $\mu^{(2)}_{\mathrm{loop}}$ and
$\mu^{(3)}_{\mathrm{loop}}$ can be calculated by using the
parameters in Eq.~\eqref{Parameters} as inputs.

\subsection{Numerical results and discussions}
The numerical results for the magnetic moments $\mu_V$ calculated in
the SU(2) and SU(3) cases are given order by order in the lower half
parts of Tables~\ref{SU2_case} and~\ref{SU3_case}, respectively. We
see that the convergence of the chiral expansion in the SU(2) case
remains very good and the convergence is also reasonable in SU(3).

In the SU(2) case, the magnetic moments at $\mathcal{O}_\mu(p^1)$
are independent of $\Delta$. The $\Delta\neq0$ correction reduces
the $\mu_V$ at $\mathcal{O}_\mu(p^2)$ and $\mathcal{O}_\mu(p^3)$.
Consequently, the total results are increased. In the heavy quark
limit, there exists a strict relationship between $\mu_V$ and
$\mu_{V\to P\gamma}$ at each order, i.e., $|\mu_V|=|\mu_{V\to
P\gamma}|$ when we take $D=4$ and $\Delta=0$ in the loop functions.
Both the radiative transitions and magnetic moments of the heavy
vector mesons are solely governed by the light quark since the heavy
quark decouples completely.

In the SU(3) case, one notices the similar variation trend at
$\mathcal{O}_\mu(p^2)$ as in SU(2). At $\mathcal{O}_\mu(p^3)$, there
is a moderate increasement when the mass splitting is included. The
total results are enhanced in the $\Delta\neq0$ case. It's
interesting to diagnose the convergence of the chiral expansion for
magnetic moments from a straightforward dimensional analysis.

The magnetic moments $\mu_V$ at the leading order (LO),
next-to-leading order (NLO), and next-to-next-to-leading order
(NNLO) can be parameterized as follows,
\begin{eqnarray}
\mathrm{LO}:\quad &&A\frac{1}{m_q}+B\frac{1}{m_Q},\nonumber\\
\mathrm{NLO}:\quad &&C\frac{m_\phi}{\Lambda_\chi^2},\nonumber\\
\mathrm{NNLO}:\quad
&&\left(D\frac{1}{m_q}+E\frac{1}{m_Q}\right)\times\frac{m_\phi^2}{\Lambda_\chi^2},
\end{eqnarray}
where the coefficients $A,\dots,E$ are order-one dimensionless
constants. $\Lambda_\chi\sim1$ GeV denotes the chiral breaking
scale.

For the $D^{\ast-}$ and $B^{\ast0}$ mesons, the internal light
pseudoscalar lines in the $\mathcal{O}(p^3)$ loop diagrams
(Figs.~\ref{Loop_Diagrams_Moments}$(a)$
and~\ref{Loop_Diagrams_Moments}$(b)$) can only be the charged pions.
But for the $\mathcal{O}(p^4)$ wave function renormalization
diagrams
(Figs.~\ref{Loop_Diagrams_Moments}$(l)$-\ref{Loop_Diagrams_Moments}$(o)$),
$K$ and $\eta$ would contribute to the loops. Since
$m_K/m_\pi\simeq3.5$ and $m_\eta/m_\pi\simeq4.0$, the
$\mathcal{O}(p^4)$ contribution would be enhanced to the same
magnitude as the $\mathcal{O}(p^3)$ correction from the SU(3)
violation effect. Let's take the $D^{\ast-}$ meson as an example. In
the strict heavy quark limit, the contributions of
Figs.~\ref{Loop_Diagrams_Moments}$(a)$
and~\ref{Loop_Diagrams_Moments}$(b)$ are equal. However, for the
charmed mesons, the mass splitting $\Delta> m_\pi$. Hence the
amplitudes of Figs.~\ref{Loop_Diagrams_Moments}$(a)$
and~\ref{Loop_Diagrams_Moments}$(b)$ are of similar size but with
opposite sign, which makes the contributions of these two diagrams
for $D^{\ast-}$ largely cancel with each other. This effect does not
contribute to the transition magnetic moments, because there is only
a single one-loop diagram with $\Delta=0$ at $\mathcal{O}(p^3)$
level (see Fig. \ref{Loop_Diagrams}$(a)$). Moreover, the influence
of the mass splitting on the magnetic properties of the $B^\ast$ is
not obvious due to $\Delta\ll m_\phi$ in the bottom sector.

The magnetic moments for the $\bar{D}^\ast$ and $B^\ast$ mesons
calculated in different cases are shown in
Table~\ref{MagneticMomentsDB}, where the errors also stem from
$\mu^{(3)}_{\mathrm{tree}}$, i.e., $\mathcal{O}(p^4)$ Lagrangians
and quark models. The magnetic moments of the vector $\bar{Q}u$,
$\bar{Q}d$ and $\bar{Q}s$ states given by the bag
model~\cite{Bose:1980vy,Simonis:2018rld} and Nambu-Jona-Lasinio
model~\cite{Luan:2015goa} are compatible with our predictions.

\begin{table*}[htbp]
\renewcommand{\arraystretch}{1.5}
 \tabcolsep=1.2pt
\caption{The magnetic moments of the charmed and bottom vector
mesons (in units of nucleon magnetons $\mu_N$), and a comparison
with the Bag model (Bag), extended Nambu-Jona-Lasinio model (NJL)
and extended Bag model (Extended Bag)
predictions.}\label{MagneticMomentsDB} \setlength{\tabcolsep}{4.5mm}
{
\begin{tabular}{c|cc|cc|ccc}
\hline\hline
\multirow{2}{*}{States}&\multicolumn{2}{c|}{SU(2)}&\multicolumn{2}{c|}{SU(3)}&\multicolumn{3}{c}{The results from other theoretical works}\\
&$\Delta=0$&$\Delta\neq0$&$\Delta=0$&$\Delta\neq0$&Bag~\cite{Bose:1980vy} &NJL~\cite{Luan:2015goa} &Extended Bag~\cite{Simonis:2018rld}\\
\hline
$\bar{D}^{\ast0}$&$1.38^{+0.25}_{-0.25}$&$1.60^{+0.25}_{-0.25}$&$1.18^{+0.25}_{-0.25}$&$1.48^{+0.22}_{-0.38}$&$0.89$&$\cdots$&$1.28$\\
$D^{\ast-}$&$-1.14_{-0.15}^{+0.15}$&$-1.39_{-0.15}^{+0.15}$&$-1.31_{-0.15}^{+0.20}$&$-1.62_{-0.08}^{+0.24}$&$-1.17$&$-1.16$&$-1.13$\\
$D_s^{\ast-}$&$\cdots$&$\cdots$&$-0.62_{-0.15}^{+0.15}$&$-0.69_{-0.10}^{+0.22}$&$-1.03$&$-0.98$&$-0.93$\\
\hline
$B^{\ast+}$&$1.86^{+0.25}_{-0.25}$&$1.90^{+0.20}_{-0.20}$&$1.71^{+0.25}_{-0.25}$&$1.77^{+0.25}_{-0.30}$&$1.54$&$1.47$&$1.56$\\
$B^{\ast0}$&$-0.75_{-0.11}^{+0.11}$&$-0.78_{-0.11}^{+0.11}$&$-0.87_{-0.11}^{+0.13}$&$-0.92_{-0.11}^{+0.15}$&$-0.64$&$\cdots$&$-0.69$\\
$B_s^{\ast0}$&$\cdots$&$\cdots$&$-0.25_{-0.11}^{+0.11}$&$-0.27_{-0.10}^{+0.13}$&$-0.47$&$\cdots$&$-0.51$\\
\hline\hline
\end{tabular}
}
\end{table*}

\begin{table*}
\renewcommand{\arraystretch}{1.5}
\caption{The transition magnetic moments and magnetic moments of the
charmed and bottom vector mesons calculated in the SU(2) case order
by order (in units of $\mu_N$).}\label{SU2_case}
\setlength{\tabcolsep}{2.5mm} {
\begin{tabular}{c|cccc|cccc}
\hline\hline
\multirow{2}{*}{Physical quantity}&\multicolumn{4}{c|}{$\Delta=0$}&\multicolumn{4}{c}{$\Delta\neq0$}\\
&$\mathcal{O}_\mu(p^1)$ Tree&$\mathcal{O}_\mu(p^2)$ Loop&$\mathcal{O}_\mu(p^3)$ Loop&Total&$\mathcal{O}_\mu(p^1)$ Tree&$\mathcal{O}_\mu(p^2)$ Loop&$\mathcal{O}_\mu(p^3)$ Loop&Total\\
\hline
$\mu_{\bar{D}^{\ast0}\to\bar{D}^0\gamma}$&$-2.24$&$0.21$&$-0.10$&$-2.13$&$-2.24$&$0.29$&$0.04$&$-1.91$\\
$\mu_{D^{\ast-}\to D^-\gamma}$&$0.55$&$-0.21$&$0.05$&$0.39$&$0.55$&$-0.29$&$0.02$&$0.28$\\
$\mu_{B^{\ast+}\to B^+\gamma}$&$-1.80$&$0.16$&$-0.09$&$-1.73$&$-1.80$&$0.19$&$-0.07$&$-1.68$\\
$\mu_{B^{\ast0}\to B^0\gamma}$&$0.99$&$-0.16$&$0.046$&$0.88$&$0.99$&$-0.19$&$0.04$&$0.84$\\
\hline
$\mu_{\bar{D}^{\ast0}}$&$1.48$&$-0.21$&$0.11$&$1.38$&$1.48$&$0.07$&$0.05$&$1.60$\\
$\mu_{D^{\ast-}}$&$-1.31$&$0.21$&$-0.05$&$-1.14$&$-1.31$&$-0.07$&$-0.007$&$-1.39$\\
$\mu_{B^{\ast+}}$&$1.93$&$-0.16$&$0.09$&$1.86$&$1.93$&$-0.13$&$0.09$&$1.90$\\
$\mu_{B^{\ast0}}$&$-0.86$&$0.16$&$-0.05$&$-0.75$&$-0.86$&$0.13$&$-0.05$&$-0.78$\\
\hline\hline
\end{tabular}
}
\end{table*}
\begin{table*}
\renewcommand{\arraystretch}{1.5}
\caption{The transition magnetic moments and magnetic moments of
charmed and bottom vector mesons calculated in the SU(3) case order
by order (in units of $\mu_N$).}\label{SU3_case}
\setlength{\tabcolsep}{2.5mm} {
\begin{tabular}{c|cccc|cccc}
\hline\hline
\multirow{2}{*}{Physical quantity}&\multicolumn{4}{c|}{$\Delta=0$}&\multicolumn{4}{c}{$\Delta\neq0$}\\
&$\mathcal{O}_\mu(p^1)$ Tree&$\mathcal{O}_\mu(p^2)$ Loop&$\mathcal{O}_\mu(p^3)$ Loop&Total&$\mathcal{O}_\mu(p^1)$ Tree&$\mathcal{O}_\mu(p^2)$ Loop&$\mathcal{O}_\mu(p^3)$ Loop&Total\\
\hline
$\mu_{\bar{D}^{\ast0}\to\bar{D}^0\gamma}$&$-2.24$&$0.71$&$-0.34$&$-1.86$&$-2.24$&$0.81$&$-0.13$&$-1.57$\\
$\mu_{D^{\ast-}\to D^-\gamma}$&$0.55$&$-0.21$&$0.19$&$0.54$&$0.55$&$-0.29$&$0.08$&$0.34$\\
$\mu_{D_s^{\ast-}\to D_s^-\gamma}$&$0.20$&$-0.50$&$0.15$&$-0.15$&$0.20$&$-0.51$&$0.10$&$-0.21$\\
$\mu_{B^{\ast+}\to B^+\gamma}$&$-1.80$&$0.55$&$-0.34$&$-1.58$&$-1.80$&$0.58$&$-0.30$&$-1.52$\\
$\mu_{B^{\ast0}\to B^0\gamma}$&$0.99$&$-0.16$&$0.17$&1.0&$0.99$&$-0.19$&$0.14$&$0.95$\\
$\mu_{B_s^{\ast0}\to B_s^0\gamma}$&$0.65$&$-0.39$&$0.13$&$0.38$&$0.65$&$-0.39$&$0.11$&$0.36$\\
\hline
$\mu_{\bar{D}^{\ast0}}$&$1.48$&$-0.71$&$0.40$&$1.18$&$1.48$&$-0.40$&$0.40$&$1.48$\\
$\mu_{D^{\ast-}}$&$-1.31$&$0.21$&$-0.21$&$-1.31$&$-1.31$&$-0.07$&$-0.24$&$-1.62$\\
$\mu_{D_s^{\ast-}}$&$-0.96$&$0.50$&$-0.16$&$-0.62$&$-0.96$&$0.47$&$-0.21$&$-0.69$\\
$\mu_{B^{\ast+}}$&$1.93$&$-0.55$&$0.34$&$1.71$&$1.93$&$-0.52$&$0.36$&$1.77$\\
$\mu_{B^{\ast0}}$&$-0.86$&$0.16$&$-0.17$&$-0.87$&$-0.86$&$0.13$&$-0.19$&$-0.92$\\
$\mu_{B_s^{\ast0}}$&$-0.51$&$0.39$&$-0.13$&$-0.25$&$-0.51$&$0.38$&$-0.14$&$-0.27$\\
\hline\hline
\end{tabular}
}
\end{table*}

\section{Summary}\label{sec5}

For the ground vector $\bar{Q}q$ states, heavy quark spin symmetry
implies the mass splitting between the spin triplets $V$ and spin
singlets $P$ is very small, which is of the same order as the pion
mass $m_\pi$. Thus the decay modes of $V$ are largely restricted.
For the ground-state charmed vector mesons, the dominant decay
channels are $V\to P\pi$ and $V\to P\gamma$. For the $b\bar{q}$
states, the only dominant decay modes are $V\to P\gamma$.

In this work, we calculate the decay rates of $V\to P\gamma$ for the
charmed and bottom vector mesons. Our result for $D^{\ast-}\to
D^-\gamma$ is in accordance with the experimental measurement. We
also investigate the convergence of the chiral expansion of the
transition magnetic moments in the SU(2) and SU(3) cases with the
mass splitting kept and unkept. The results indicate that the
convergence in SU(2) case is very good, and it is reasonable for
SU(3) likewise. The effect of the mass splitting for the charmed
mesons is more significant than that for the bottom mesons. The
radiative decay widths of the $D^\ast$ and $B^\ast$ mesons from
other theoretical models and lattice QCD simulations also are
consistent with ours. As a byproduct, the full widths of
$\bar{D}^{\ast0}$ and $D_s^{\ast-}$ are estimated to be
$77.7^{+26.7}_{-20.5}$ keV and $0.62^{+0.45}_{-0.50}$ keV,
respectively.

In this work, we also calculate the magnetic moments of the $D^\ast$
and $B^\ast$ mesons. Our results agree with the predictions of bag
model~\cite{Bose:1980vy,Simonis:2018rld} and NJL
model~\cite{Luan:2015goa}. The magnetic moments of heavy vector
mesons are good platforms to probe their inner structures. For
example, the magnetic moment of $\bar{D}^{\ast0}$ should be zero if
we use the classical formula
$\boldsymbol{\mu}=\frac{e}{2m}\mathbf{S}$ (where $e$, $m$ and
$\mathbf{S}$ denote the charge, mass and spin, respectively).
The large anomalous magnetic moment of $\bar{D}^{\ast0}$ clearly
demonstrates that it is not a point particle.

In summary, we have systematically studied the radiative transitions
and magnetic moments of charmed and bottom vector mesons with
$\chi$PT up to $\mathcal{O}(p^4)$. Our numerical results are
presented up to this order with different scenarios. The LECs
$\tilde{a}$, $a$ and $b$ in the $\mathcal{O}(p^2)$ Lagrangians are
estimated with the quark model and resonance saturation model,
respectively. We notice the one-loop chiral correction plays a very
important role in mediating the (transition) magnetic moments. Our result
indicates the quark model prediction is not enough to describe the
magnetic properties of the charmed and bottom vector mesons.
The quark dynamics of the light degree of freedom that is related with
the spontaneous breaking of chiral symmetry is non-negligible.

The present investigations of the radiative decays of $D^\ast$ and
$B^\ast$ shall be helpful to the future measurement at facilities
such as BelleII and LHCb. Furthermore, the analytical expressions
derived in $\chi$PT shall be helpful for the chiral extrapolations
of lattice QCD simulations on the electromagnetic transitions and
magnetic moments of heavy vector mesons.

\section*{Acknowledgments}
B. W is very grateful to X. L. Chen and W. Z. Deng for helpful
discussions. This project is supported by the National Natural
Science Foundation of China under Grants 11575008, 11621131001 and
National Key Basic Research Program of China(2015CB856700).

\appendix
\section{Some supplemental materials for the $B^\ast$ mesons}\label{appdixA}
The transition form factors from Figs.~\ref{Tree_Diagrams}$(a)$
and~\ref{Tree_Diagrams}$(b)$ for the $B^\ast$ mesons read
\begin{eqnarray}
\mu^{\prime(a)}_{B^{\ast+}\to B^+\gamma}&=&\frac{8}{3}(3a+2\tilde{a}),\\
\mu^{\prime(a)}_{B^{\ast0}\to B^0\gamma}&=&\frac{8}{3}(3a-\tilde{a}),\\
\mu^{\prime(a)}_{B_s^{\ast0}\to B_s^0\gamma}&=&\frac{8}{3}(3a-\tilde{a}),\\
\mu^{\prime(b)}_{B^{\ast+}\to B^+\gamma}&=&-\frac{32}{9}(m_K^2-m_\pi^2)(3\tilde{d}+3\bar{d}+4d),\\
\mu^{\prime(b)}_{B^{\ast0}\to B^0\gamma}&=&-\frac{32}{9}(m_K^2-m_\pi^2)(3\tilde{d}+3\bar{d}-2d),\\
\mu^{\prime(b)}_{B_s^{\ast0}\to
B_s^0\gamma}&=&-\frac{32}{9}(m_K^2-m_\pi^2)(-6\tilde{d}+3\bar{d}+4d).
\end{eqnarray}

The magnetic moments from Figs.~\ref{Tree_Diagrams_Moments}$(a)$
and~\ref{Tree_Diagrams_Moments}$(b)$ for the $B^\ast$ mesons read
\begin{eqnarray}
\mu_{B^{\ast+}}^{(a)}&=&\frac{4}{3}e(-2\tilde{a}+3a),\\
\mu_{B^{\ast0}}^{(a)}&=&\frac{4}{3}e(\tilde{a}+3a),\\
\mu_{B_s^{\ast0}}^{(a)}&=&\frac{8}{3}e(\tilde{a}+3a),\\
\mu_{B^{\ast+}}^{(b)}&=&\frac{16}{9}e(m_K^2-m_\pi^2)(-3\tilde{d}+3\bar{d}+4d),\\
\mu_{B^{\ast0}}^{(b)}&=&\frac{16}{9}e(m_K^2-m_\pi^2)(-3\tilde{d}+3\bar{d}-2d),\\
\mu_{B_s^{\ast0}}^{(b)}&=&\frac{16}{9}e(m_K^2-m_\pi^2)(6\tilde{d}+3\bar{d}+4d).
\end{eqnarray}

The flavor dependent coefficients $\mathcal{C}_\phi^{(x)}$ in
Eqs.~\eqref{CphiAD}-\eqref{CphiIJD} and
Eqs.~\eqref{CphiADast}-\eqref{CphilmDast} for the $B^\ast$ mesons
are listed in Tables~\ref{Flavor_Coe_B1}-\ref{Flavor_Coe_B2} and
Tables~\ref{Flavor_Coe_Bast1}-\ref{Flavor_Coe_Bast2}, respectively.
\section{Loop integrals}\label{appdixB}
Here, we show the detailed forms of the $\mathcal{J}$ functions used
in the text. One can find the complete forms in Ref.
\cite{Wang:2018atz}.
\begin{widetext}
\begin{eqnarray}
i\int\frac{d^Dl\lambda^{4-D}}{(2\pi)^D}\frac{1}{l^2-m^2+i\varepsilon}&\equiv&\mathcal{J}_0^c(m),\label{LoopIntJc}\\
i\int\frac{d^Dl\lambda^{4-D}}{(2\pi)^D}\frac{l^\alpha l^\beta}{\left(v\cdot l+\omega+i\varepsilon\right)\left(l^2-m^2+i\varepsilon\right)}&\equiv& \left[v^\alpha v^\beta \mathcal{J}_{21}^a+g^{\alpha\beta}\mathcal{J}_{22}^a\right](m,\omega),\label{LoopIntJa}\\
i\int \frac{d^Dl\lambda^{4-D}}{(2\pi)^D}\frac{l^\alpha
l^\beta}{\left(v\cdot l+\omega+i\varepsilon\right)\left[v\cdot
l+\delta+i\varepsilon\right]\left(l^2-m^2+i\varepsilon\right)}
&\equiv&\left[v^\alpha v^\beta \mathcal{J}_{21}^g+ g^{\alpha\beta}\mathcal{J}_{22}^g\right](m,\omega,\delta),\label{LoopIntJgh}\\
i\int\frac{d^Dl\lambda^{4-D}}{(2\pi)^D}\frac{l^\alpha l^\beta}{\left(l^2-m^2+i\varepsilon\right)\left[(l+q)^2-m^2+i\varepsilon\right]}&\equiv&\left[q^\alpha q^\beta \mathcal{J}_{21}^F+g^{\alpha\beta}\mathcal{J}_{22}^F\right](m,q),\label{LoopIntJF}\\
i\int\frac{d^Dl\lambda^{4-D}}{(2\pi)^D}\frac{l^\alpha l^\beta}{\left(v\cdot l+\omega+i\varepsilon\right)\left(l^2-m^2+i\varepsilon\right)\left[(l+q)^2-m^2+i\varepsilon\right]}&\equiv&\left[g^{\alpha\beta}\mathcal{J}_{21}^T+q^\alpha q^\beta \mathcal{J}_{22}^T+v^\alpha v^\beta \mathcal{J}_{23}^T+(q\vee v)\mathcal{J}_{24}^T\right](m,\omega,q),\label{LoopIntJT}\nonumber\\
\end{eqnarray}
\end{widetext}
where $q\vee v\equiv q^\alpha v^\beta+q^\beta v^\alpha$. The
$\mathcal{J}$ functions defined above can be calculated with the
dimensional regularization in $D$ dimensions. In the following, we
write out the expressions of the used $\mathcal{J}$ functions.
\begin{widetext}
\begin{eqnarray}
\mathcal{J}_0^c(m)&=&2m^2L+\frac{m^2}{16\pi^2}\ln\frac{m^2}{\lambda^2},\label{LoopIntJ0c}\\
\mathcal{J}_{22}^a(m,\omega)&=&2\omega\left(m^2-\frac{2}{3}\omega^2\right)L+\frac{1}{16\pi^2}\int_{-\omega}^0\tilde{\Delta}\ln\frac{\tilde{\Delta}}{\lambda^2}dy+\frac{1}{24\pi}\tilde{A}^{3/2},\label{LoopIntJ22a}\\
\mathcal{J}_{22}^g(m,\omega,\delta)&=&\left\{\begin{array}{ll}
\frac{1}{\delta-\omega}\left[\mathcal{J}_{22}^a(m,\omega)-\mathcal{J}_{22}^a(m,\delta)\right] & \textrm{if $\omega\neq\delta$}\\
-\frac{\partial}{\partial
x}\mathcal{J}_{22}^a(m,x)\Big|_{x\to\omega(\textrm{or}\ \delta)} &
\textrm{if $\omega=\delta$}
\end{array} \right.,\\
\mathcal{J}_{22}^F(m,q)&=&\left(m^2-\frac{q^2}{6}\right)L+\frac{1}{32\pi^2}\int_0^1\bar{\Delta}\ln\frac{\bar{\Delta}}{\lambda^2}dx,\label{LoopIntJ22F}\\
\mathcal{J}_{21}^T(m,\omega,q)&=&2\omega
L+\frac{1}{16\pi^2}\int_0^1dx\int_{-\omega}^0\left(1+\ln\frac{\Delta}{\lambda^2}\right)dy+\frac{1}{16\pi}\int_0^1A^{1/2}dx,
\end{eqnarray}
where
$\tilde{\Delta}=y^2+\tilde{A},~\tilde{A}=m^2-\omega^2-i\varepsilon$;
$\bar{\Delta}=x(x-1)q^2+m^2-i\varepsilon$; $\Delta=y^2+A$,
$A=x(x-1)q^2+m^2-(\omega-xq_0)^2-i\varepsilon$.
\end{widetext}

$L$ is defined as
\begin{eqnarray}
L=\frac{1}{16\pi^2}\left[\frac{1}{D-4}+\frac{1}{2}\left(\gamma_E-1-\ln4\pi\right)\right],
\end{eqnarray}
where $\gamma_E$ is the Euler-Mascheroni constant $0.5772157$. We
adopt the $\overline{\mathrm{MS}}$ scheme to renormalize the loop
integrals, which is equivalent to making use of the following
relation,
\begin{eqnarray}
\{X\}_r=\lim_{D\to4}\left(X-L\frac{\partial}{\partial
L}X\right)+\frac{1}{16\pi^2}\lim_{D\to4}\left(\frac{\partial}{\partial
D}\frac{\partial}{\partial L}X\right),
\end{eqnarray}
where $\{X\}_r$ represents the finite part of $X$.
\section{Estimating the light quark mass with vector meson dominance model}\label{appdixC}
In general, the transition form factor of $V\to P\gamma$ at the leading order can be parameterized as follows,
\begin{eqnarray}
\mu_{\bar{Q}q}^\prime=\mathcal{Q}_{\bar{Q}}\frac{1}{\Lambda_{\bar{Q}}}-\mathcal{Q}_q\frac{1}{\Lambda_q},
\end{eqnarray}
where $\mathcal{Q}_{\bar{Q}}$ and $\mathcal{Q}_q$ denote the charge matrices of $\bar{Q}$ and $q$, respectively. $\Lambda_{\bar{Q}}$ and $\Lambda_q$ are
the mass parameters that can be understood as the masses of the constituent quarks in the quark model. Heavy quark symmetry guarantees $\Lambda_{\bar{Q}}\approx m_{\bar{Q}}$ (see the discussions in Ref. \cite{Manohar:2000dt}).
However, the photon coupling to the light quark part of the electromagnetic current is not fixed by the heavy quark symmetry, thus the $\Lambda_q$ is not a ``well-defined" constant, its value is largely model dependent to some extent. Here, we adopt the vector meson dominance (VMD) model \cite{Pacetti:2015iqa,Colangelo:1993zq} to estimate the value of $\Lambda_q$.

\begin{figure}[hptb]
%\vspace{0.3cm}
    %\setlength{\abovecaptionskip}{0.1cm}
    %\setlength{\belowcaptionskip}{-0.4cm}
    \scalebox{1.0}{\includegraphics[width=\columnwidth]{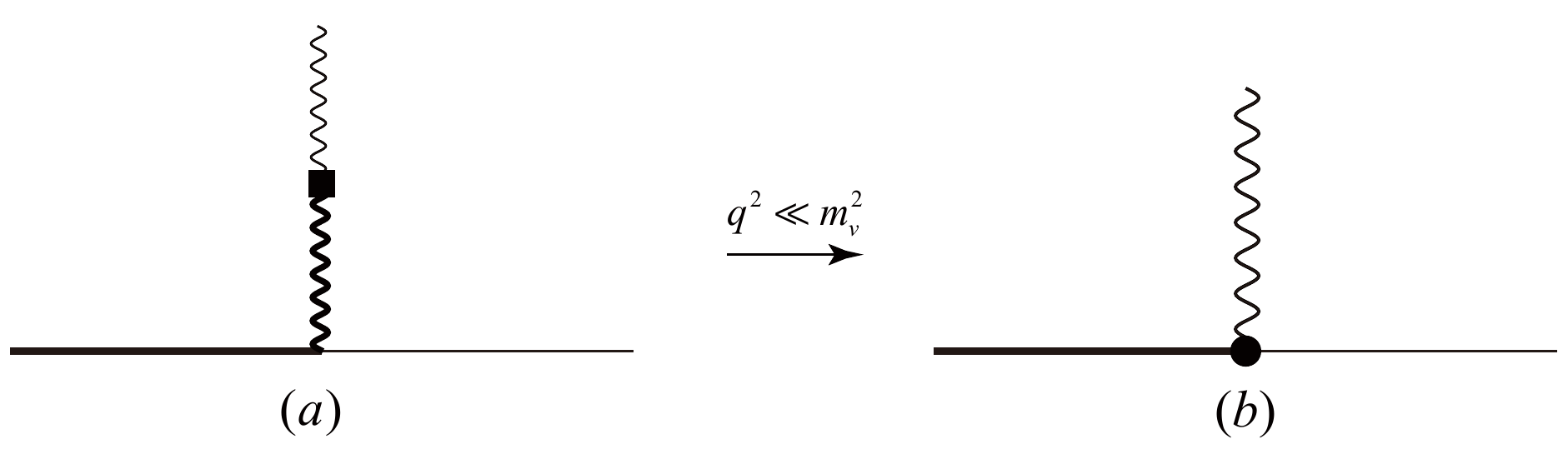}}
    \caption{A diagrammatic presentation of the vector meson dominance model. The thick wiggly line in figure $(a)$ denotes the light vector meson $\rho$ or $\phi$, the solid square denotes the coupling vertex of photon and light vector meson. Other notations are same as those in Fig.~\ref{Loop_Diagrams}.\label{VMD}}
\end{figure}

In the VMD model, the light quark part of the electromagnetic current $\langle P|J_\mu^\ell|V\rangle$ can be expressed as follows by inserting the light vector resonance $\mathcal{V}$,
\begin{eqnarray}\label{DDVgamma}
\langle P_a(p^\prime)|J_\mu^\ell(q^2)|V_a(p,\varepsilon_V)\rangle
&=&ie_a\sum_{\mathcal{V},\lambda}\frac{\langle0|\bar{q}_a\gamma_\mu q_a|\mathcal{V}(q,\varepsilon_{\mathcal{V}}^\lambda)\rangle}{q^2-m_{\mathcal{V}}^2}\nonumber\\
&&\times\langle P_a(p^\prime)\mathcal{V}(q,\varepsilon_{\mathcal{V}}^\lambda)|V_a(p,\varepsilon_V)\rangle,\nonumber\\
\end{eqnarray}
where the $\langle P_a\mathcal{V}|V_a\rangle$ vertex is given in Eq. \eqref{LHrho} (A diagrammatic presentation of Eq. \eqref{DDVgamma} is shown in Fig. \ref{VMD}). The matrix element $\langle0|\bar{q}_a\gamma_\mu q_a|\mathcal{V}(q,\varepsilon_{\mathcal{V}}^\lambda)\rangle$ can be calculated by assuming the SU(3) symmetry with
\begin{eqnarray}
\langle0|\bar{q}_a\gamma_\mu q_a|\mathcal{V}(q,\varepsilon_{\mathcal{V}}^\lambda)\rangle=f_{\mathcal{V}}\varepsilon_{\mathcal{V}}^\mu\mathrm{Tr}(\mathcal{V}T^a),
\end{eqnarray}
where $f_{\mathcal{V}}$ and $\varepsilon_{\mathcal{V}}$ denote the decay constant and polarization vector of the light vector meson, respectively. $(T^a)_{lm}=\delta_{al}\delta_{am}$, and $a=1,2,3$ for $u,d,s$, respectively. The $f_{\mathcal{V}}$ can be determined by the electromagnetic decay $\mathcal{V}\to e^+e^-$. $f_\rho=0.17$ GeV$^2$ for the $\rho$ meson, and $f_\phi=0.25$ GeV$^2$ for the $\phi$ meson \cite{Colangelo:1993zq}.

Following the same procedure in obtaining Eq. \eqref{aat}, one can get
\begin{eqnarray}\label{Lambdaq}
\Lambda_q^{-1}=2\sqrt{2}g_v\lambda\sqrt{\frac{m_V}{m_P}}\frac{f_\mathcal{V}}{m_\mathcal{V}^2},
\end{eqnarray}
where the values of $g_v$ and $\lambda$ are the same as those in Eq. \eqref{bvalue}. One can obtain $\Lambda_q$ by considering the SU(3) breaking effect in Eq. \eqref{Lambdaq}, eventually,
\begin{eqnarray}
\Lambda_u=\Lambda_d=0.366 ~\mathrm{GeV}, \quad \Lambda_s=0.596 ~\mathrm{GeV}.
\end{eqnarray}
These values are very close to the $m_u$, $m_d$ and $m_s$ given in Eq. \eqref{Parameters}.
\begin{table*}[!htbp]
\renewcommand{\arraystretch}{1.7}
 \tabcolsep=1.8pt
\caption{The flavor-dependent coefficients
$\mathcal{C}_\phi^{(x)}~(x=a,\dots,d)$ in
Eqs.~\eqref{CphiAD}-\eqref{CphiDD} for the $B^\ast$
mesons.}\label{Flavor_Coe_B1} \setlength{\tabcolsep}{4.2mm} {
\begin{tabular}{c|ccccccccc}
\hline\hline
Decay modes& $\mathcal{C}_\pi^{(a)}$ &$\mathcal{C}_K^{(a)}$ &$\mathcal{C}_\pi^{(b)}$ &$\mathcal{C}_K^{(b)}$ &$\mathcal{C}_\pi^{(c)}$ &$\mathcal{C}_K^{(c)}$&$\mathcal{C}_\pi^{(d)}$&$\mathcal{C}_K^{(d)}$&$\mathcal{C}_\eta^{(d)}$\\
\hline
$B^{\ast+}\to B^+\gamma$&$2$&$2$&$-4$&$-4$&$4$&$4$&$-12a$&$-\frac{8}{3}(3a+\tilde{a})$&$-\frac{4}{9}(3a-2\tilde{a})$\\
$B^{\ast0}\to B^0\gamma$&$-2$&$0$&$4$&$0$&$-4$&$0$&$-4(3a-\tilde{a})$&$-\frac{8}{3}(3a+\tilde{a})$&$-\frac{4}{9}(3a+\tilde{a})$\\
$B^{\ast0}_s\to B^0_s\gamma$&$0$&$-2$&$0$&$4$&$0$&$-4$&$0$&$-\frac{8}{3}(6a-\tilde{a})$&$-\frac{16}{9}(3a+\tilde{a})$\\
\hline\hline
\end{tabular}
}
\end{table*}
\begin{table*}[!htbp]
\renewcommand{\arraystretch}{1.7}
 \tabcolsep=1.2pt
\caption{The flavor-dependent coefficients
$\mathcal{C}_\phi^{(x)}~(x=e,\dots,j)$ in
Eqs.~\eqref{CphiED}-\eqref{CphiIJD} for the $B^\ast$
mesons.}\label{Flavor_Coe_B2} \setlength{\tabcolsep}{5.25mm} {
\begin{tabular}{c|cccccc}
\hline\hline
Decay modes& $\mathcal{C}_\pi^{(e)}$ &$\mathcal{C}_K^{(e)}$ &$\mathcal{C}_\eta^{(e)}$ &$\mathcal{C}_\pi^{(h)}$ &$\mathcal{C}_K^{(h)}$ &$\mathcal{C}_\eta^{(h)}$\\
\hline
$B^{\ast+}\to B^+\gamma$&$-6a$&$-\frac{4}{3}(3a-\tilde{a})$&$-\frac{2}{9}(3a+2\tilde{a})$&$3a+2\tilde{a}$&$\frac{2}{3}(3a+2\tilde{a})$&$\frac{1}{9}(3a+2\tilde{a})$\\
$B^{\ast0}\to B^0\gamma$&$-2(3a+\tilde{a})$&$-\frac{4}{3}(3a-\tilde{a})$&$-\frac{2}{9}(3a-\tilde{a})$&$3a-\tilde{a}$&$\frac{2}{3}(3a-\tilde{a})$&$\frac{1}{9}(3a-\tilde{a})$\\
$B^{\ast0}_s\to B^0_s\gamma$&$0$&$-\frac{4}{3}(6a+\tilde{a})$&$-\frac{8}{9}(3a-\tilde{a})$&$0$&$\frac{4}{3}(3a-\tilde{a})$&$\frac{4}{9}(3a-\tilde{a})$\\
\hline\hline
\end{tabular}
}
\end{table*}
\begin{table*}[!htbp]
\renewcommand{\arraystretch}{1.7}
 \tabcolsep=1.8pt
\caption{The flavor-dependent coefficients
$\mathcal{C}_\phi^{(x)}~(x=a,c,d,e)$ in
Eqs.~\eqref{CphiADast}-\eqref{CphiEDast} for the $B^\ast$
mesons.}\label{Flavor_Coe_Bast1} \setlength{\tabcolsep}{4.7mm} {
\begin{tabular}{c|ccccccccc}
\hline\hline
States& $\mathcal{C}_\pi^{(a)}$ &$\mathcal{C}_K^{(a)}$ &$\mathcal{C}_\pi^{(c)}$ &$\mathcal{C}_K^{(c)}$ &$\mathcal{C}_\pi^{(d)}$ &$\mathcal{C}_K^{(d)}$&$\mathcal{C}_\pi^{(e)}$&$\mathcal{C}_K^{(e)}$&$\mathcal{C}_\eta^{(e)}$\\
\hline
$B^{\ast+}$&$-\frac{1}{2}$&$-\frac{1}{2}$&$2$&$2$&$-2$&$-2$&$-3a$&$-\frac{2}{3}(3a-\tilde{a})$&$-\frac{1}{9}(3a+2\tilde{a})$\\
$B^{\ast0}$&$\frac{1}{2}$&$0$&$-2$&$0$&$2$&$0$&$-(3a+\tilde{a})$&$-\frac{2}{3}(3a-\tilde{a})$&$-\frac{1}{9}(3a-\tilde{a})$\\
$B_s^{\ast0}$&$0$&$\frac{1}{2}$&$0$&$-2$&$0$&$2$&$0$&$-\frac{2}{3}(6a+\tilde{a})$&$-\frac{4}{9}(3a-\tilde{a})$\\
\hline\hline
\end{tabular}
}
\end{table*}
\begin{table*}[!htbp]
\renewcommand{\arraystretch}{1.7}
 \tabcolsep=1.2pt
\caption{The flavor-dependent coefficients
$\mathcal{C}_\phi^{(x)}~(x=g,l+m)$ in
Eqs.~\eqref{CphiGDast}-\eqref{CphilmDast} for the $B^\ast$
mesons.}\label{Flavor_Coe_Bast2} \setlength{\tabcolsep}{5.52mm} {
\begin{tabular}{c|cccccc}
\hline\hline
States& $\mathcal{C}_\pi^{(g)}$ &$\mathcal{C}_K^{(g)}$ &$\mathcal{C}_\eta^{(g)}$ &$\mathcal{C}_\pi^{(lm)}$ &$\mathcal{C}_K^{(lm)}$ &$\mathcal{C}_\eta^{(lm)}$\\
\hline
$B^{\ast+}$&$-3a$&$-\frac{2}{3}(3a+\tilde{a})$&$-\frac{1}{9}(3a-2\tilde{a})$&$\frac{1}{2}(2\tilde{a}-3a)$&$\frac{1}{3}(2\tilde{a}-3a)$&$\frac{1}{18}(2\tilde{a}-3a)$\\
$B^{\ast0}$&$-(3a-\tilde{a})$&$-\frac{2}{3}(3a+\tilde{a})$&$-\frac{1}{9}(3a+\tilde{a})$&$-\frac{1}{2}(3a+\tilde{a})$&$-\frac{1}{3}(3a+\tilde{a})$&$-\frac{1}{18}(3a+\tilde{a})$\\
$B_s^{\ast0}$&$0$&$-\frac{2}{3}(6a-\tilde{a})$&$-\frac{4}{9}(3a+\tilde{a})$&$0$&$-\frac{2}{3}(3a+\tilde{a})$&$-\frac{2}{9}(3a+\tilde{a})$\\
\hline\hline
\end{tabular}
}
\end{table*}

\end{document}